\documentclass[aps,prb,twocolumn,superscriptaddress,preprintnumbers]{revtex4-2}

\usepackage{amsmath, amssymb}
\usepackage{xcolor} 
\usepackage{soul}
\usepackage{float}
\usepackage[colorlinks = true, urlcolor = cyan, linkcolor=blue, citecolor=blue, filecolor=magenta]{hyperref}
\usepackage{adjustbox}
\usepackage{booktabs}
\usepackage{physics}

\newcommand{\Zd}{\mathbb{Z}_d}
\newcommand{\Xd}{X_d}
\newcommand{\Zdop}{Z_d}
\newcommand{\Wpq}{W_{p,q}}
\newcommand{\CSUM}{\mathtt{CSUM}}
\newcommand{\CSUB}{\mathtt{CSUB}}
\newcommand{\Uenc}{U_{\mathrm{enc}}}
\newcommand{\Udec}{U_{\mathrm{dec}}}

\newcommand{\Sb}{S_B}
\newcommand{\id}{\mathbb{I}}

\newcommand{\figbox}[2]{%
  \IfFileExists{#1}{\includegraphics[width=#2]{#1}}%
  {\fbox{\parbox[c][50mm][c]{#2}{\centering\ttfamily\detokenize{#1}\\[4pt]
   \normalfont\itshape figure to be supplied}}}}

\usepackage{xcolor}

\begin{document}

\title{Secret-shared teleportation of a two-qudit gate with a deferred recipient}

\author{Tanay Roy}
\email{roytanay@fnal.gov}
\affiliation{Superconducting Quantum Materials and Systems (SQMS) Center, Fermi National Accelerator Laboratory, Batavia, IL 60510, USA}

\preprint{FERMILAB-PUB-26-0765-SQMS}

\date{\today}

\begin{abstract}
An unknown quantum operation, like an unknown quantum state, cannot be perfectly copied or stored for later use. A specific gate can be embedded in an entangled resource and teleported, but its destination is fixed when the resource is created. Here we show that the capability to apply a two-qudit entangling gate can instead be secret-shared among many parties: the shares are distributed in advance, and a later key release lets exactly one of them, chosen after distribution, exercise it. A provider encrypts half of a Bell pair held with a user into $n$ shares, one per candidate node. The selected node, given the key qudits, completes an exact non-local $\CSUM^{k}$ gate with the user's data, in every measurement branch and for any qudit dimension, with the power $k$ chosen by the user alone. Any coalition of fewer than $n$ candidate nodes, or the key register alone, holds no information, unselected nodes cannot tell whether a selection occurred, and the key cannot decrypt twice. For two candidates the registers form a $((3,5))$ threshold scheme. Encoder and decoder are explicit $O(n)$ circuits. The construction supports jointly authorized and anonymous gate application, advancing distributed quantum computing and quantum network capabilities.
\end{abstract}

\maketitle

Quantum information cannot be copied~\cite{Wootters1982no_cloning_qubit},
and neither can an unknown quantum operation: no finite quantum program
can store an arbitrary unitary for later
execution~\cite{NielsenChuang1997programmable}, and copies of a gate are
necessarily imperfect~\cite{Chiribella2008no_cloning_gate}. What can be
stored is a \emph{specific} gate, frozen into an entangled resource and
released later by local measurement and classical
feedforward~\cite{GottesmanChuang1999teleport,Eisert2000,Huelga2001remote}.
Gate teleportation of this kind is the elementary operation of
distributed quantum computing and of a quantum
internet~\cite{kimble2008internet,wehner2018internet}:
it applies a non-local entangling gate between data held at different nodes
without moving the data, and it has been demonstrated between logical
qubits, between separated trapped ions and across optical network
links~\cite{Chou2018teleport,Wan2019teleport,Main2025distributed}.
In every such protocol, however, the resource is bound to its
destination when it is created. The half of the entangled pair that
completes the gate is sent to one node, and whoever holds it holds the
gate.

Secret sharing offers a different way to hold a resource. A threshold
scheme splits a secret among $n$ parties so that only authorized subsets
recover it and all other sets learn nothing~\cite{Shamir1979share}. Its
quantum analog shares an unknown quantum state in the same
way~\cite{Hillery1999sharing,Cleve1999share,Karlsson1999sharing},
with constructions based on graph states and on absolutely maximally
entangled (AME) states~\cite{Keet2010qudit,Helwig2012AME}, states that are maximally entangled across every bipartition,
and demonstrations with photons and optical modes, recently in eleven
dimensions~\cite{Bell2014graph,Pinnell2020elevendim}.
In all of these, the shared object is a state. Authority over an
operation has also been shared, in quantum operation
sharing~\cite{Zhang2011shared}, controlled remote implementation of
operations~\cite{Wang2007controlled} and threshold-controlled
teleportation~\cite{SaiToh2009threshold}. There, however, the recipient
is fixed in advance, and every controller must measure and announce an
outcome, so each can veto the operation and all of them observe it.
Here, we ask: can the \emph{capability to act}, namely, applying an entangling
gate to data that never enter the scheme, be secret-shared so that the node exercising it is chosen after distribution and the
unselected sharers take no part?

A gate shared in this way would meet three requirements of an
entanglement-based network: joint authorization, anonymity and early
distribution. Authority to use a gate can be split among independent
custodians, none of whom individually holds anything of value, as
threshold cryptography does for classical keys. Moreover, nodes that are not
chosen should learn neither the data nor whether anyone was chosen, a
requirement shared with anonymous quantum
communication~\cite{Christandl2005anonymous} and with
secure multiparty and delegated quantum
computation~\cite{Crepeau2002multiparty,Broadbent2009blind}. Because
remote entanglement is slow to establish~\cite{Main2025distributed},
the user's entanglement should be in place before the destination is
known. However, the
obvious ways of deferring the destination fail at least one of these
requirements. For example, a provider that shares a Bell pair with the user and
later swaps its half onto the chosen node defers the choice, but the
chooser already holds the gate. A Greenberger--Horne--Zeilinger (GHZ) state
spread over the user and all candidates, as in quantum information
splitting~\cite{Hillery1999sharing,Karlsson1999sharing}, requires every
unselected node to measure and announce its outcome, so any one of them
can veto the gate, and it leaves each share classically correlated with
the user (Supplementary Section~\ref{si:context}).

\begin{figure*}[t]
    \centering
    \includegraphics[width=\textwidth]{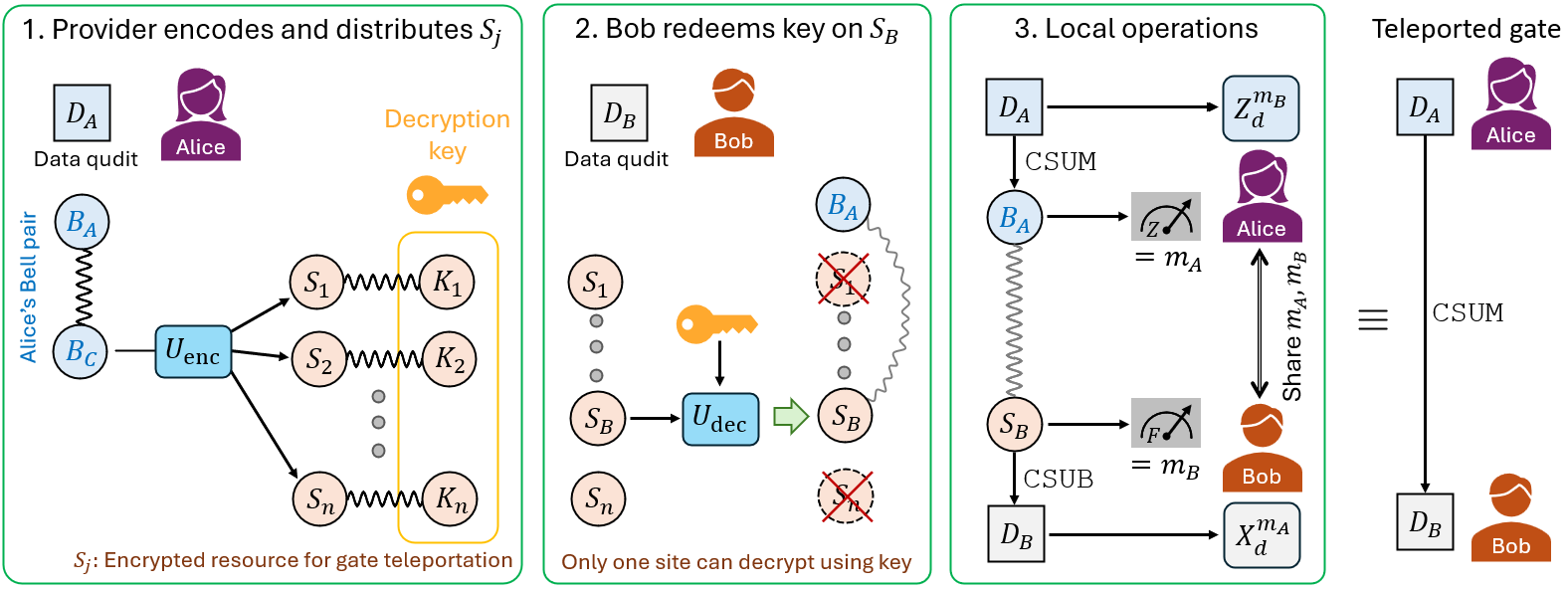}
    \caption{\textbf{Deferred-recipient teleportation of a two-qudit gate.}
In stage 1, a maximally entangled pair $\ket{\Phi}_{B_AB_C}$ is shared
between Alice, who holds $B_A$, and the provider, who holds $B_C$; it
may be prepared by either party. The provider also prepares $n$ Bell
pairs $\ket{\Phi}_{S_jK_j}$ and applies the encrypted-cloning encoder
$\Uenc$ to $B_C$ and the signal qudits $S_1,\dots,S_n$, then sends one
signal to each of $n$ candidate sites and withholds the key qudits
$K_1,\dots,K_n$. Each share alone carries no information, and the
provider keeps nothing that can complete the gate. Alice's data qudit
$D_A$ is untouched at this stage.
The recipient (Bob) is chosen only after distribution in the next stage:
releasing the key register to one site lets it apply $\Udec$ and recover
$B_C$, with its entanglement with $B_A$, at unit fidelity. All other
shares (dashed, crossed) stay maximally mixed, their holders cannot tell
that a selection was made, and the spent key cannot decrypt them.
In the final stage, Alice applies $\CSUM(D_A\!\to\!B_A)$ and Bob applies
$\CSUB(S_B\!\to\!D_B)$. One computational-basis and one Fourier-basis
measurement, two publicly exchanged dits and the Weyl corrections
$X_d^{m_A}$ on $D_B$ and $Z_d^{m_B}$ on $D_A$ complete an exact
$\CSUM(D_A\!\to\!D_B)$ in every measurement branch, although $D_A$ and
$D_B$ never interact directly. Replacing the two gates by $\CSUM^{k_1}$
and $\CSUB^{k_2}$ yields $\CSUM^{k_1k_2}$.}
    \label{fig:fig1}
\end{figure*}

Encrypted cloning~\cite{YamaguchiKempf2026cloning} supplies the missing
ingredient. It splits a qubit into $n$ signal copies, each
indistinguishable from noise, any one of which can be decrypted once by
whoever holds an entangled key. Qudit versions of encrypted cloning have recently been developed~\cite{Ceara2026qudit,LimLo2026AME}, with Ref.~\cite{LimLo2026AME}
relating the construction to AME states and secret sharing. Here we
apply encrypted cloning not to data but to one half of a
gate-teleportation resource, and show that the result is a threshold
quantum secret-sharing scheme for a non-local two-qudit gate. Specifically, a
provider distributes encrypted shares to $n$ candidate nodes. Releasing
$n$ key qudits later lets exactly one node complete an exact $\CSUM$
with the user, in every measurement branch, for any dimension $d$ and
for data entangled with arbitrary external systems, while every
coalition of fewer than $n$ candidates, or the key register alone,
holds nothing. We derive the access structure of this encoder in
closed form; for two candidates the five distributed registers form a
$((3,5))$ threshold scheme, as anticipated in Ref.~\cite{LimLo2026AME}. The user alone decides which power $\CSUM^{k}$ is
applied, and $k$ is not revealed in any transmitted register or public message.
Encoder and decoder are explicit circuits of $O(n)$
gates, independent of $d$, and every claim is proved in closed form.
We work with qudits because high-dimensional carriers are increasingly
available~\cite{Ringbauer2022qudit,kim2026ultracoherent,huang2026sideband} and bring higher capacity
and noise tolerance to cryptographic tasks~\cite{Cerf2002dlevel}. The
construction is novel already for $d=2$; for $d>2$ it relies on two
features of the arithmetic of the qudit Clifford
group~\cite{Gottesman1999,Appleby2005clifford} that are not present for qubits: a flat phase angle that depends on the
parity of $d$, which we show is necessary for the quadratic-phase
encoder, and sign bookkeeping in
the teleportation stage that makes every power of $\CSUM$ available.

\section*{Protocol}

\begin{figure*}[t]
\centering
\includegraphics[width=\textwidth]{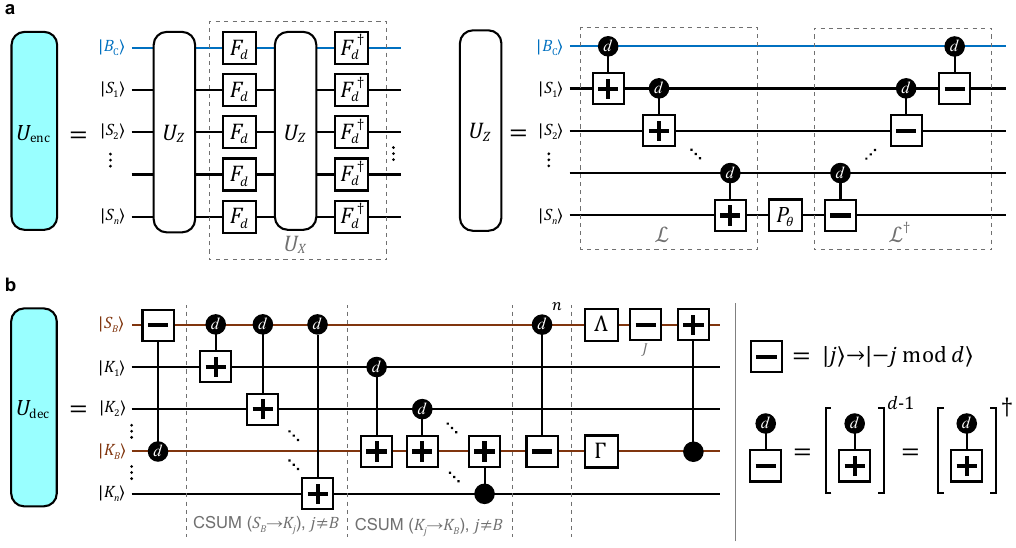}
\caption{\textbf{Explicit encoding and decoding circuits.}
\textbf{a}, The encoder $U_{\rm enc}$ of Eq.~\eqref{eq:encoder}: a $\CSUM$ ladder
accumulates the total register sum onto $S_n$, a single quadratic phase
gate $P_{\theta^{\ast}}$ is applied, and the ladder is inverted; the
whole block is then repeated conjugated by $F^{\otimes(n+1)}$. Key
registers $K_j$ are spectators throughout. \textbf{b}, The decoder
$U_{\rm dec}^{(B)}$ of Eq.~\eqref{eq:decoder}, acting only on
$(S_B,K_1,\dots,K_n)$: $\CSUB(K_B\!\to\!S_B)$ extracts the shift label
onto $S_B$; $\CSUM(S_B\!\to\!K_j)$ restores each unselected pair to
$\ket\Phi$; $\CSUM(K_j\!\to\!K_B)$ and $\CSUB(S_B\!\to\!K_B)^{n}$
collect the phase label onto $K_B$; the single-qudit gates $\Lambda$
and $\Gamma$ cancel the two Gauss-sum chirps; and $J$ followed by
$\CSUM(K_B\!\to\!S_B)$ recombines. Total: $3n$ two-qudit and three
single-qudit gates, independent of the $d^{2(n+1)}$ Hilbert-space
dimension a generic recovery map would act on.}
\label{fig:circuits}
\end{figure*}

The protocol runs in three stages as depicted in Fig.~\ref{fig:fig1}. 

\textbf{Stage 1 --- distributing the resource.}
A provider, who may be Alice herself, prepares a
maximally entangled pair $\ket\Phi_{B_AB_C}$ and gives $B_A$ to Alice. By applying $U_{\rm enc}$
it encodes $B_C$ into $n$ signal qudits $S_1,\dots,S_n$, each initially
Bell-paired with a key qudit $K_j$, sends $S_j$ to candidate site $j$
and retains the key register $(K_1,\dots,K_n)$. Alice's data qudit
$D_A$ plays no part and need not yet exist. 

\textbf{Stage 2 --- one site redeems the key.}
At any
later time and by any criterion external to the protocol, the provider
selects a site, Bob, and releases the key register to that site alone.
Bob applies a decoder $\Udec$ to $(\Sb,K_1,\dots,K_n)$ and recovers
$B_C$ with its full entanglement with $B_A$, at fidelity exactly one:
$(B_A,\Sb)$ is then a perfect Bell pair between Alice and Bob. Every
other share is left untouched and maximally mixed.
\emph{Operational requirement:} the key register must reach Bob intact
over an authenticated quantum channel~\cite{Barnum2002authentication};
the keys alone carry no information, but an unselected site that
obtained them could redeem the gate in Bob's place.

\textbf{Stage 3 --- local operations complete the gate.}
Alice and Bob complete a
non-local gate by local operations and one round of classical
communication. Alice applies $\CSUM(D_A\!\to\!B_A)$ and measures $B_A$
in the computational basis (outcome $m_A$); Bob applies
$\CSUB(\Sb\!\to\!D_B)$ and measures $\Sb$ in the Fourier basis
$\{F^{\dagger}\ket m\}$, i.e.\ applies $F$ (defined below) and reads
out $m_B$ in the computational basis; after exchanging the two dits,
Bob applies $\Xd^{m_A}$ to
$D_B$ and Alice applies $\Zdop^{m_B}$ to $D_A$. Only Stage~1 requires
quantum communication with all candidates, and the two dits of Stage~3
may be announced publicly (Methods).
\emph{Operational requirement:} the dits need integrity but no
secrecy~\cite{WegmanCarter1981}; altering them by $(\delta_A,\delta_B)$
replaces $\CSUM$ by $(\Zdop^{\delta_B}\otimes\Xd^{\delta_A})\CSUM$, a
definite Weyl error rather than a randomization (Methods).

\noindent\textbf{Trust assumptions and anonymity.} The provider is trusted to
follow the protocol and knows the recipient, as does Alice. Candidate
sites are colluding semi-honest: they may store, measure and pool shares
but do not deviate. Anonymity refers to these unselected sites: their
registers and the public transcript $(m_A,m_B)$ are independent of
whether, and which, site was selected (Supplementary
Section~\ref{si:context}).

\section*{Encoding and decoding circuits}

The construction uses the shift and clock operators
$\Xd\ket{j}=\ket{j+1}$ and $\Zdop\ket{j}=\omega^{j}\ket{j}$ with
$\omega=e^{2\pi i/d}$ and all labels taken modulo $d$; the discrete
Fourier transform $F_{kj}=\omega^{jk}/\sqrt{d}$; the sum gate
$\CSUM\ket{c,t}=\ket{c,c+t}$ and its inverse $\CSUB$; and the quadratic
phase gate
\begin{equation}
  P_\theta=\mathrm{diag}\!\left(e^{i\theta k^{2}}\right)_{k=0}^{d-1}.
  \label{eq:phasegate}
\end{equation}
With $\mathcal{L}$ the ladder
$\CSUM(B_C\!\to\!S_1),\CSUM(S_1\!\to\!S_2),\dots,\CSUM(S_{n-1}\!\to\!S_n)$
and $\mathbb{F}=F^{\otimes(n+1)}$ on $(B_C,S_1,\dots,S_n)$, the encoder
is
\begin{equation}
  \Uenc=U_XU_Z,\quad
  U_Z=\mathcal{L}^{\dagger}\left(P_\theta\ \text{on}\ S_n\right)\mathcal{L}, \quad
  U_X=\mathbb{F}^{\dagger}U_Z\,\mathbb{F},
  \label{eq:encoder}
\end{equation}
which costs $4n$ two-qudit gates, two phase gates and two Fourier
layers (Fig.~\ref{fig:circuits}a). For $d=2$ it reduces to the qubit
encoder of Ref.~\cite{YamaguchiKempf2026cloning}; the Pauli-exponential
form used there has no qudit counterpart, which is why we build the
encoder as a circuit (Methods). Each $S_j$ starts in the Bell state
$\ket\Phi_{S_jK_j}=d^{-1/2}\sum_k\ket{kk}$. Privacy of the unselected
shares holds for every $\theta$, but exact recovery requires $\theta$
to be a flat angle (below), the smallest of which is
\begin{equation}
  \theta^{\ast}(d)=\frac{2\pi}{d\,\gcd(2,d)}
  =\begin{cases}2\pi/d, & d\ \text{odd},\\[2pt] \pi/d, & d\ \text{even}.\end{cases}
  \label{eq:critical}
\end{equation}

The decoder $U_{\rm dec}$ acts only on $(\Sb,K_1,\dots,K_n)$ and never touches $B_A$,
$D_A$, $D_B$ or any unselected share. It is an explicit circuit,
obtained by inverting the Weyl-sum form of the encoder term by term
(Methods). With $\nu(d)=\gcd(2,d)$, $\nu_{\rm inv}=2/\nu$ (always
invertible modulo $d$) and
\begin{equation}
  \lambda(d)=\begin{cases}\omega^{4^{-1}}, & d\ \text{odd},\\[2pt]
                          e^{i\pi/d},      & d\ \text{even},\end{cases}
  \label{eq:lambda}
\end{equation}
where $4^{-1}$ is the inverse of $4$ modulo $d$, it is the ordered
product, read right to left,
\begin{equation}
\begin{aligned}
  \Udec^{(B)}=\ &\CSUM(K_B\!\to\!S_B)\cdot J_{S_B}\cdot
                 \Gamma_{K_B}\cdot\Lambda_{S_B}\\
   &\cdot\CSUB(S_B\!\to\!K_B)^{\,n} \cdot
    \prod_{j\ne B}\CSUM(K_j\!\to\!K_B)\\
   &\cdot\prod_{j\ne B}\CSUM(S_B\!\to\!K_j)\cdot\CSUB(K_B\!\to\!S_B),
\end{aligned}
  \label{eq:decoder}
\end{equation}
with single-qudit gates $\Lambda=\mathrm{diag}(\lambda^{k^{2}})$,
$J\ket{x}=\ket{-x}$ and
$\Gamma=P_{\theta^{\ast}}^{\dagger}R\,F^{\dagger}P_{\theta^{\ast}}^{\dagger}$,
where $R\ket{u}=\ket{\nu_{\rm inv}^{-1}u}$. The decoder uses $3n$
two-qudit and three single-qudit gates (Fig.~\ref{fig:circuits}b),
independent of $d$.

\begin{figure*}[t]
\centering
\figbox{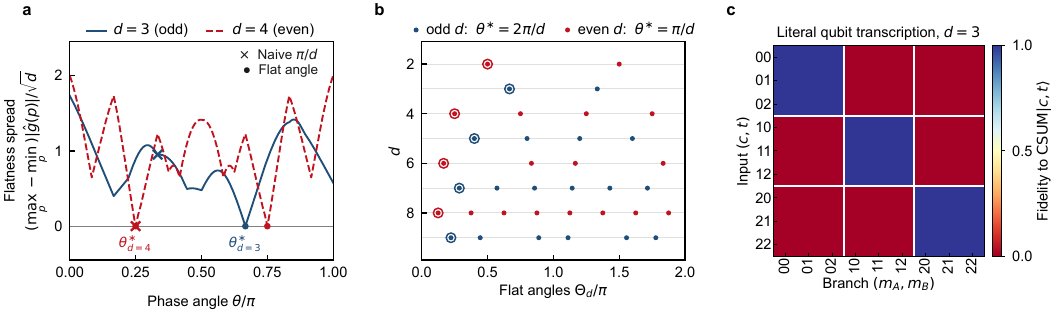}{\textwidth}
\caption{\textbf{Two features hidden at $d=2$.}
\textbf{a}, The flatness spread $\max_p|\hat g(p)|-\min_p|\hat g(p)|$,
in units of $\sqrt d$, as a function of the phase angle $\theta$ for one
odd ($d=3$) and one even ($d=4$) dimension; filled circles mark its
exact zeros. The naive rescaling of the qubit angle $\pi/2$ to $\pi/d$
($\times$) is flat for $d=4$ but not for $d=3$, where
$|\hat g(p)|=(1,\sqrt7,1)$ and the spread is $(\sqrt7-1)/\sqrt3\approx
0.95$. \textbf{b}, The zero set is the lattice
$\Theta_d=\{\pi s/d:\gcd(s,d)=1,\ sd\ \text{even}\}$, shown for
$d=2,\dots,9$, with open rings on its smallest positive member
$\theta^{\ast}$: $2\pi/d$ for odd $d$ and $\pi/d$ for even $d$.
\textbf{c}, Fidelity to $\CSUM\ket{c,t}$ at $d=3$ of the literal
transcription of the qubit circuit, in which both parties apply
$\CSUM$ and the qubit corrections $\Xd^{m_A}$, $\Zdop^{m_B}$ are used,
for every input $(c,t)$ and every measurement branch $(m_A,m_B)$. This
circuit teleports $\CSUB$, so it reproduces $\CSUM$ only on the
$\nu(d)/d$ fraction with $m_A\equiv c\ (\mathrm{mod}\ d/\nu(d))$ (the
diagonal blocks), a coincidence with the unmeasurable logical value
$c$. Choosing opposite signs for the two gates, $k_1k_2\equiv1$ in
Eq.~\eqref{eq:csumk}, makes every entry equal to 1 (not shown).}
\label{fig:obstruct}
\end{figure*}

\section*{Exact gate, privacy and single use}

The central result is that the protocol implements the gate exactly,
for any input and in every measurement branch. Take any $d\ge2$ and
$n\ge2$, an encoder with $\theta=\theta^{\ast}$ (or any flat angle, with
the decoder adjusted as in Methods), and any site $B$. Let Alice apply $\CSUM^{k_1}(D_A\!\to\!B_A)$ and Bob
$\CSUB^{k_2}(\Sb\!\to\!D_B)$, with corrections $\Xd^{k_2m_A}$ on $D_B$
and $\Zdop^{k_1m_B}$ on $D_A$, and set $k\equiv k_1k_2\pmod d$
(Fig.~\ref{fig:fig1} is $k_1=k_2=1$). The input is an arbitrary
density operator $\rho$ on $D_AD_BR$, where $R$ is a reference system
of any dimension, held by anyone, on which the protocol does not act.
With ideal operations, delivery of the intact key register to $B$ and
faithful exchange of $(m_A,m_B)$, every branch $(m_A,m_B)$ occurs with
probability $1/d^{2}$, independently of $\rho$, and the corrected state
conditioned on that branch is
\begin{equation}
  \rho_{\rm out}^{(m_A,m_B)}
  =\big(\CSUM^{k}\otimes\id_R\big)\,\rho\,
   \big(\CSUM^{k}\otimes\id_R\big)^{\dagger},
  \label{eq:main}
\end{equation}
with $\CSUM^{k}$ acting as $D_A\!\to\!D_B$. All other registers end in
a state that does not depend on $\rho$ and is uncorrelated with
$D_AD_BR$.

For a pure input
$\ket{\chi}=\sum_{a,b}c_{a,b}\ket{a}_{D_A}\ket{b}_{D_B}$ and $k=1$, Eq.~\eqref{eq:main} reads
$\CSUM\ket{\chi}=\sum_{a,b}c_{a,b}\,\ket{a}_{D_A}\ket{b+a}_{D_B}$.
Eq.~\eqref{eq:main} equates branch-conditioned outputs, not branch
averages: nothing is post-selected, every branch is corrected to the
same target, and because $R$ is arbitrary it covers mixed inputs and
inputs entangled with systems outside anyone's control. The derivation
(Methods) combines two exact identities: the decoder returns a perfect
Bell pair between Alice and Bob, Eq.~\eqref{eq:recovery}, and that pair
teleports $\CSUM^{k}$ deterministically, Eq.~\eqref{eq:teleport}.

Privacy is equally exact. Any set of unselected shares---one site, or
all $n-1$ non-recipients pooling their qudits---is left in
$\id/d^{|\mathcal C|}\otimes\rho_E$, maximally mixed and uncorrelated
with any external system $E$, for every $d$, every $n\ge2$ and every
phase angle (Methods). An unselected site therefore learns nothing
about Alice's data and cannot distinguish a run in which a different
site was selected from one in which no site was selected at all. The
key is single-use. After decoding, the full register is the product of
maximally entangled pairs
\begin{equation}
  \ket{\Phi}_{B_A S_B}\otimes\ket{\Phi}_{B_C K_B}
  \otimes\bigotimes_{j\ne B}\ket{\Phi}_{S_j K_j},
  \label{eq:postdecode}
\end{equation}
in which every unselected share is re-paired with its own key and
retains no correlation with $B_A$. A second site that applies the
decoder with the spent key obtains entanglement fidelity $1/d^{2}$ with
$B_A$, the value of a random guess. Only the coalition of all $n$
candidates, which before selection includes the future recipient, can
be correlated with $B_A$. It too is private if $n$ is chosen with
$\gcd(n-\eta^{2},d)=1$, where $\eta=1$ for even $d$ and $\eta=2^{-1}$
modulo $d$ for odd $d$; for $n=2$ this holds unless $7\mid d$
(Supplementary Section~\ref{si:coalition}).

\section*{Secret sharing of a gate}

The encoded register is a quantum secret-sharing scheme. The secret is
$B_C$'s entanglement with $B_A$, and the $2n+1$ shareholders are
$B_C$, the $n$ signal qudits and the $n$ key qudits. At a flat angle
the encoder is a Clifford circuit, so the encoded register is a
stabilizer state, and its access structure follows exactly from
counting solutions of linear congruences modulo $d$ (Supplementary
Section~\ref{si:access}). A set of registers that misses at least one
signal--key pair entirely holds the secret if and only if it contains
$B_C$ and at least one complete pair $(S_j,K_j)$; a set that touches
every pair holds it if it contains any complete pair; and the only
remaining sets, which hold exactly one member of each pair, carry an
amount of information fixed by a single gcd condition extending
earlier classifications of informative subsets in encrypted
cloning~\cite{Gianini2026leak,Gianini2026full,Bai2026classification}.

Three consequences give the scheme its operational meaning. First, the
key register on its own carries no information for any $d$ and $n$.
Handing the $n$ key qudits to $n$ independent custodians therefore
makes the gate available only if every custodian releases its key,
while the custodians together, without a share, learn nothing---joint
authorization of a quantum operation rather than of a message. Second,
the recipient's set $(\Sb,K_1,\dots,K_n)$ is authorized without $B_C$,
so the provider may discard $B_C$ once the shares are sent and never
itself holds the gate. Third, for $n=2$ the five registers form a
$((3,5))$ threshold scheme, the qudit analog of the scheme that Ref.~\cite{Cleve1999share} obtains from the five-qubit code: any three
recover the secret, and any two may be lost or intercepted without
leaking anything or destroying the gate. Lim and Lo showed that
encrypted cloning with two pairs is equivalent to such a scheme in
every dimension~\cite{LimLo2026AME}; for the present encoder the
access structure shows that this holds at $\theta^{\ast}$ whenever
$\gcd(7,d)=1$, and for $d=7$ another flat angle restores it. For $n\ge3$ the scheme is no longer threshold:
some unauthorized sets hold exactly half of the secret, as in ramp
schemes~\cite{Ogawa2005ramp}. Unlike in secret sharing of a state,
the reconstructed secret is an entangled resource that the recipient
spends on a gate with data that never entered the scheme: the protocol
shares a \emph{gate}.

\section*{Two features of qudit arithmetic}

Two elements of the construction are invisible at $d=2$ and must be
handled explicitly for $d>2$ (Fig.~\ref{fig:obstruct}). Both reflect known features of the qudit
Clifford group~\cite{Gottesman1999,Appleby2005clifford}.
The first is a genuine constraint: for the quadratic-phase encoder of
Eq.~\eqref{eq:encoder}, we show that no other angle gives exact
recovery by any decoder. The second is a bookkeeping rule that, once made
explicit, turns into a resource (Methods).

\textbf{The flat angle splits by parity.}
Because $B_C$ never reaches the recipient, exact recovery requires
the provider's pair to be left maximally mixed, which forces the Gauss
sum $\hat g(p)=\sum_k e^{i\theta k^2}\omega^{-pk}$ to be flat,
$|\hat g(p)|=\sqrt d$ for every $p$ (Methods). This holds exactly on the lattice
$\theta=\pi s/d$ with $\gcd(s,d)=1$ and $sd$ even (see Fig.~\ref{fig:obstruct}b), whose smallest
positive member is Eq.~\eqref{eq:critical}. For odd $d$ it is reached
by completing the square with the inverse of $2$ modulo $d$; for even
$d$ that inverse does not exist, and a $2d$-th root of unity is needed
instead. Rescaling the qubit angle $\pi/2$ to $\pi/d$ therefore fails
for every odd $d$; at $d=3$ it gives $|\hat g(p)|=(1,\sqrt7,1)$
(Fig.~\ref{fig:obstruct}a,b).

\textbf{The teleported power is a product of local powers.}
If Alice applies $\CSUM^{k_1}(D_A\!\to\!B_A)$ and Bob applies
$\CSUB^{k_2}(\Sb\!\to\!D_B)$, then with corrections $\Xd^{k_2m_A}$ on
$D_B$ and $\Zdop^{k_1m_B}$ on $D_A$ the protocol applies
\begin{equation}
  \CSUM^{k}(D_A\!\to\!D_B),\qquad k\equiv k_1k_2 \pmod d,
  \label{eq:csumk}
\end{equation}
exactly in every branch; the protocol of Fig.~\ref{fig:fig1} is
$k_1=k_2=1$. Because $\CSUM$ is its own inverse only for qubits, the literal transcription of the qubit circuit, in which
both parties apply $\CSUM$ ($k_1=1$, $k_2=-1$), teleports $\CSUB$; used
with the qubit corrections it leaves $D_B$ displaced by an amount that
depends on the unknown logical value of $D_A$
(Fig.~\ref{fig:obstruct}c). Any choice with $k_1k_2\equiv1$ gives
$\CSUM$; which party carries the inverse is immaterial. More usefully,
Eq.~\eqref{eq:csumk} makes every power of $\CSUM$ reachable with one
Bell pair and two dits. If Bob fixes $k_2=1$, Alice alone chooses
$k=k_1$, possibly after the recipient has been selected. Bob's
operations, the key register and the public dits $(m_A,m_B)$, which are
uniformly distributed for every $k$, are then all independent of $k$:
the protocol reveals nothing about the power applied beyond what the
gate's own action on Bob's data qudit can reveal (Methods).

\begin{figure*}[t]
\centering
\figbox{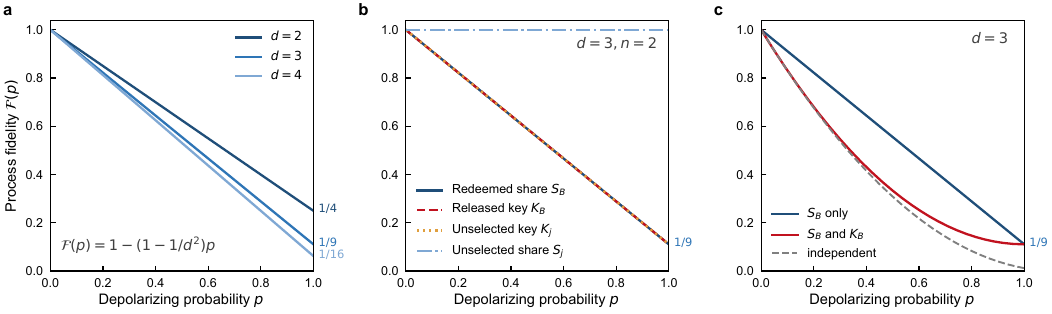}{\textwidth}
\caption{\textbf{Performance under depolarizing noise.}
Each panel is computed exactly: the channel
$\mathcal D_p(\rho)=(1-p)\rho+p\,\mathbb I\Tr[\rho]/d$ is expanded in
Weyl operators and every branch propagated as a pure state, so no
sampling is involved. The figure of merit is the process fidelity of
the completed non-local $\CSUM$, with $D_A$ and $D_B$ each maximally
entangled with a reference qudit, summed over all $d^{2}$ measurement
branches. \textbf{a}, Process fidelity versus the depolarizing
probability $p$ on the redeemed share $S_B$, for $d=2,3,4$ at $n=2$. The
response is exactly linear, Eq.~\eqref{eq:noiselinear}, from $1$ at
$p=0$ to $1/d^{2}$ at $p=1$. \textbf{b}, The same, resolved by which
register is exposed ($d=3$, $n=2$). Noise on the redeemed share $S_B$,
on the released key $K_B$ and on an unselected key $K_j$ act
identically; noise on an unselected share $S_j$ leaves the gate
untouched. \textbf{c}, The floor does not compound for errors on $S_B$ and $K_B$:
depolarizing both (red) keeps the $1/d^{2}$ floor rather than falling
to the $1/d^{4}$ that independent errors would give (gray dashed).}
\label{fig:noise}
\end{figure*}

\section*{Effect of noise on the distributed registers}

To move beyond ideal operations we apply the qudit depolarizing
channel $\mathcal D_p(\rho)=(1-p)\rho+p\,\mathbb I\Tr[\rho]/d$ to the
qudits that travel or are stored between encoding and decoding, and
compute the process fidelity $\mathcal{F}$ of the completed gate exactly (Methods).
The model is deliberately limited to this storage and transmission
stage: state preparation, the encoder and decoder gates, the Stage~3
operations and both measurements are taken to be ideal. The results
therefore quantify how errors on the distributed registers propagate
to the gate. They are not an end-to-end fault-tolerance analysis,
which would require encoding the full register (Discussion). Noise on
the redeemed share ($S_B$) gives
\begin{equation}
  \mathcal{F}(p)=1-\Big(1-\frac{1}{d^{2}}\Big)p ,
  \label{eq:noiselinear}
\end{equation}
linear from $1$ to the random-unitary value $1/d^{2}$, with no
threshold (Fig.~\ref{fig:noise}a). Noise on the unselected shares has
no effect at all, whereas the key qudits are as sensitive as the
redeemed share (Fig.~\ref{fig:noise}b; for $K_B$ this holds whenever
$\gcd(n-\eta^{2},d)=1$), so only the redeemed share and the
key register need protection. Three aspects of the noise response
should be kept apart. The \emph{floor} does not compound: depolarizing
both $\Sb$ and $K_B$ completely still leaves $1/d^{2}$ rather than the
$1/d^{4}$ that independent errors would give (Fig.~\ref{fig:noise}c),
because part of the joint errors cancel in the decoder. The
\emph{full $p$-dependence} does compound: with both registers noisy,
$\mathcal F(p)=1-(1-1/d^{2})[1-(1-p)^{2}]$, so the infidelity roughly
doubles at small $p$. And the \emph{scaling with $n$} is set by the
number of exposed key qudits. Since $\mathcal F$ is multilinear in the
channels, its first-order term is additive, and with $S_B$ and all
$n$ keys each depolarized with probability $p$,
$1-\mathcal F\simeq(n+1)(1-1/d^{2})p$ to first order in $p$ (slightly less
when $\gcd(n-\eta^{2},d)>1$; Methods). Protecting the
key register is thus a cost that grows linearly with the number of
candidates, even though no single error is worse than the loss of one
link.

\section*{Discussion}

This protocol can be viewed as encrypted cloning of a gate rather than of a state. While an unknown unitary cannot be copied or programmed~\cite{Chiribella2008no_cloning_gate,NielsenChuang1997programmable}, our scheme offers the ability to apply one fixed gate to be distributed among $n$ sites, with exactly one site able to exercise it at a later time. For every unselected site, the scheme functions as a private quantum channel~\cite{Ambainis2000private}. It is closely related to quantum one-time programs~\cite{Broadbent2013}, copy
protection~\cite{Aaronson2009copy} and uncloneable
encryption~\cite{BroadbentLord2020uncloneable}, which typically
require computational or hardware assumptions. In contrast, our protocol achieves single use and privacy through information-theoretic means, though it is limited to a single Clifford gate and assumes parties do not deviate from the protocol. Furthermore, it extends the concept of teleporting a unitary operation~\cite{Huelga2001remote,Reznik2002remote} by allowing the destination of the operation, as well as its timing, to be decided after the resource is created. Because the power $k$ of the 
$\CSUM$ is chosen locally
by the user, the provider, the candidates and the public channel remain
ignorant of which gate was applied, a limited form of the operation
hiding sought in blind and delegated quantum
computation~\cite{Broadbent2009blind}.

The construction yields two results of independent interest. First, removing the data qudits and the teleportation stage yields qudit encrypted cloning with an explicit $3n$-gate decoder, exact transfer of entanglement with an arbitrary external system, and flat angles that are both necessary and sufficient. This complements recent qudit constructions whose decoders scale with $d$ or are given as operator sums ~\cite{Ceara2026qudit,LimLo2026AME}. Second, deleting the cloning stage
leaves deterministic non-local $\CSUM^{k}$ for every $k\in\Zd$
(Supplementary Sections~\ref{si:cloning} and~\ref{si:teleport}). The
decoder structure also has practical consequences, discussed in
Methods: all $n$ auxiliary Bell pairs survive decoding and can be
reused, a lost key can be replaced by a declined share, and Alice may
apply her gate at the last moment.

Our privacy guarantee applies to parties that observe without deviating from the protocol. However, if an unselected site obtains the entire key register, it holds an authorized set, so the key must be transmitted over an authenticated quantum channel. Active attacks on the key or on the classical channel are not covered by this protocol (Supplementary Section~\ref{si:context}). Extending the scheme to verifiable secret sharing~\cite{Crepeau2002multiparty} against active adversaries, to non-Clifford or general two-qudit gates, and to fault-tolerant encodings of the full register are natural directions for future work. The resource requirements are modest—$2n+4$ qudits, $O(n)$ two-qudit gates, and one round of classical communication—and are within reach of current qudit processors~\cite{Ringbauer2022qudit, bornman2025benchmarking, huang2026sideband} and photonic platforms where high-dimensional secret sharing has already been demonstrated~\cite{Pinnell2020elevendim}.

\section*{Methods}

\subsection*{Qudit conventions}
For a qudit with dimension $d\ge2$ and computational basis $\{\ket0,\dots,\ket{d-1}\}$, the
Weyl--Heisenberg (generalized Pauli) operators~\cite{Gottesman1999} are $\Xd\ket j=\ket{j+1
\bmod d}$ and $\Zdop\ket j=\omega^j\ket j$, which satisfy
\begin{equation}
  \Zdop\Xd=\omega\Xd\Zdop,\qquad
  \Zdop^a\Xd^b=\omega^{ab}\Xd^b\Zdop^a \quad(a,b\in\Zd),
  \label{eq:comm}
\end{equation}
where $\omega=e^{2\pi i/d}$. For $p,q\in\Zd$, $\Wpq:=\Xd^{p}\Zdop^{q}$; the $d^2$ operators
$\{\Wpq\}$ form an orthogonal operator basis,
\begin{equation}
  \Tr\!\left[W_{p,q}^{\dagger}W_{p',q'}\right]=d\,\delta_{pp'}\delta_{qq'}.
  \label{eq:orthogonality}
\end{equation}
The discrete Fourier transform $F$ ($F_{kj}=\omega^{jk}/\sqrt d$)
satisfies $F^{-1}\Zdop F=\Xd$, and the maximally entangled state
$\ket\Phi=d^{-1/2}\sum_j\ket{jj}$ obeys, for any operator $M$,
$\bra\Phi(M\otimes\mathbb{I})\ket\Phi=\Tr[M]/d$. Together with $\CSUM$, $\CSUB$
and $P_\theta$ as defined in the main text, these are the only objects
the constructions below require; all arithmetic on qudit labels is
modulo $d$ throughout.

\subsection*{Why the encoder is built as a circuit}
The qubit encrypted-cloning encoder~\cite{YamaguchiKempf2026cloning},
\begin{equation}
U_{\mathrm{enc}}^{(\rm qubit)}
= e^{-i\frac{\pi}{4}X^{(B_C)}\otimes\bigotimes_{j=1}^{n} X^{(S_j)}}\,
e^{-i\frac{\pi}{4}Z^{(B_C)}\otimes\bigotimes_{j=1}^{n} Z^{(S_j)}},
\end{equation}
is a product of two exponentials of Pauli strings, a form that relies on
$X^2=Z^2=\id_2$ and on the single-qubit Pauli group closing on four
operators. Neither fact survives for $d>2$: $\Xd$ and $\Zdop$ have
order $d$, and $\{\Wpq\}$ is a set of $d^{2}$ distinct operators, so the
two-exponential form has nothing to generalize to. We therefore define
the encoder by the circuit of Eq.~\eqref{eq:encoder} and derive its
properties from the operator structure below (Supplementary
Section~\ref{si:obstruction}).

\subsection*{Encoder construction and its Weyl-sum closed form}
The component $U_Z$ in Eq.~\eqref{eq:encoder} is diagonal in the
computational basis of $(B_C,S_1,\dots,S_n)$. This is because the forward $\CSUM$ chain $\mathcal{L}$ accumulates the running sum onto
$S_n$, the phase gate $P_\theta$ multiplies by $e^{i\theta T^2}$ with
$T=(a+s_1+\cdots+s_n)\bmod d$, and the reverse chain restores every
register, so $U_Z\ket{a,s_1,\dots,s_n}=e^{i\theta T^2}\ket{a,s_1,\dots,s_n}$.
Writing $g(k)=e^{i\theta k^2}$ in terms of its discrete Fourier
coefficients $\hat g(p)=\sum_k g(k)\,\omega^{-pk}$ (because $g(k)=g(-k)$,
$\hat g(p)=\hat g(-p)$, so the opposite sign convention gives identical
coefficients) gives $U_Z=\tfrac1d
\sum_p\hat g(p)\,(\Zdop^{p})^{\otimes(n+1)}$, and conjugating by
$\mathbb{F}=F^{\otimes(n+1)}$ turns each $\Zdop^{p}$ into $\Xd^{p}$.
Multiplying the two factors and reordering with Eq.~\eqref{eq:comm}
collects the result into the Weyl basis,
\begin{equation}
  \Uenc=\frac{1}{d^2}\sum_{p,q=0}^{d-1}\mu_{p,q}\,W_{p,q}^{\otimes(n+1)},
  \quad
  \mu_{p,q}:=\hat g(p)\,\hat g(q),
  \label{eq:weylsum}
\end{equation}
where $W_{p,q}^{\otimes(n+1)}$ acts as $\Wpq=\Xd^{p}\Zdop^{q}$ on each of
$B_C,S_1,\dots,S_n$ and as the identity on every key qudit $K_j$; with
this ordering of $\Xd$ and $\Zdop$ the reordering phases generated by
Eq.~\eqref{eq:comm} cancel identically and no residual factor survives.
Everything that follows --- the flat angle, single-share privacy,
and the explicit decoder --- is read off this one closed-form expansion.
For $d=2$ it reduces to the qubit encoder of
Ref.~\cite{YamaguchiKempf2026cloning}, up to the global phase $i$
(checked numerically for $n\le3$).

\subsection*{Flat angles and the chirp identity}
Tracing out every signal--key pair in Eq.~\eqref{eq:branches} below
(the pair states $(\Wpq\otimes\mathbb I)\ket\Phi$ are orthonormal in
$(p,q)$) leaves the provider's pair in the Bell-diagonal state
$\rho_{B_AB_C}=\sum_{p,q}\frac{|\mu_{p,q}|^{2}}{d^{4}}\,
(\Wpq)_{B_C}\ketbra{\Phi}(\Wpq)^{\dagger}_{B_C}$. The decoder never
touches $B_C$, so exact recovery of $B_A$'s partner at $S_B$ requires
$B_A$ to be decoupled from $B_C$, i.e.\ $\rho_{B_AB_C}=\id/d^{2}$, i.e.\
all $d^{2}$ weights equal. Recovery therefore requires $|\hat g(p)|$ to
be independent of $p$, which by Parseval's identity
($\sum_p|\hat g(p)|^2=d\sum_k|g(k)|^2=d^2$) forces $|\hat g(p)|=\sqrt d$
for every $p$. The scope of this necessity statement is precise: for
the quadratic-phase encoder $\Uenc(\theta)$ of Eq.~\eqref{eq:encoder},
exact recovery by \emph{any} decoder acting on
$(S_B,K_1,\dots,K_n)$ implies flatness of $|\hat g|$, which pins
$\theta$ to the set $\Theta_d$ derived below. It does not exclude
encoders outside this one-parameter family. (Privacy, by contrast,
needs no condition on $\theta$; see below.) Flatness is not an approximate
numerical fact to be checked dimension by dimension: at
$\theta^{\ast}(d)$ the Gauss sum $\hat g(p)$ is a \emph{pure chirp},
\begin{equation}
  \hat g(p)=G\,\lambda^{-p^{2}},\qquad
  G=\sum_{k\in\Zd}\lambda^{k^{2}},\qquad |G|=\sqrt d ,
  \label{eq:chirp}
\end{equation}
with $\lambda$ as in Eq.~\eqref{eq:lambda}, so that $|\hat g(p)|=|G|=
\sqrt d$ identically, for every $p$ and every $d$. The proof is
completion of the square. For odd $d$, $\theta^{\ast}=2\pi/d$ gives
$g(k)=\omega^{k^{2}}$ and $2$ is invertible modulo $d$, so
$k^{2}-pk=(k-p\cdot2^{-1})^{2}-p^{2}\cdot4^{-1}$ and re-indexing
$k\mapsto k+p\cdot2^{-1}$ yields Eq.~\eqref{eq:chirp} with
$\lambda=\omega^{4^{-1}}$. For even $d$ no such inverse exists;
instead, at $\lambda=\zeta=e^{i\pi/d}$ (a $2d$-th root of unity with
$\zeta^{2}=\omega$) one has $\hat g(p)=\sum_k\zeta^{k^{2}-2pk}=
\zeta^{-p^{2}}\sum_k\zeta^{(k-p)^{2}}$, and $\zeta^{(k+d)^{2}}=
\zeta^{k^{2}}\omega^{kd}e^{i\pi d}=\zeta^{k^{2}}$ because $d$ is even,
so the shifted sum is the unshifted one and Eq.~\eqref{eq:chirp}
follows at $\theta^{\ast}=\pi/d$, half the odd-$d$ value. In both cases
$|G|=\sqrt d$ is the classical evaluation of the quadratic Gauss sum.
The identity Eq.~\eqref{eq:chirp} is the single structural fact the
decoder inverts: all of the $p$-dependence of $\hat g$ sits in a
quadratic phase, which one single-qudit diagonal gate removes.

The same argument identifies every flat angle, not just the smallest.
Writing $\theta=\pi s/d$, the periodicity step requires
$e^{i\pi s d}=1$, i.e.\ $sd$ even, and the resulting Gauss sum
$G_s=\sum_k e^{i\pi sk^{2}/d}$ has modulus $\sqrt d$ exactly when
$\gcd(s,d)=1$; when $\gcd(s,d)=m>1$ the sum degenerates and
$|\hat g(p)|$ takes the value $\sqrt{md}$ on a sublattice of $p$ and
zero elsewhere. Angles off the lattice $\pi\Zd/d$ are irrational
multiples of the relevant root of unity and never flat. Hence the flat
set is exactly $\{\theta=\pi s/d:\gcd(s,d)=1,\ sd\ \text{even}\}$ and
$\theta^{\ast}(d)$ of Eq.~\eqref{eq:critical} is its smallest positive
member. At every flat angle completing the square again gives a pure
chirp, $\hat g(p)=G_s\lambda_s^{-p^{2}}$, now with
\begin{equation}
  \lambda_s=\begin{cases}\omega^{(2s)^{-1}}, & d\ \text{odd},\\[2pt]
                          e^{i\pi\bar s/d}, & d\ \text{even},\end{cases}
  \label{eq:lambdas}
\end{equation}
where $(2s)^{-1}$ is taken modulo $d$ and $\bar s$ is the inverse of $s$
modulo $2d$; at $\theta^{\ast}$ this reduces to Eq.~\eqref{eq:lambda}.
All flat angles are therefore equivalent for the protocol: the decoder
is unchanged except that $\Lambda=\mathrm{diag}(\lambda_s^{k^{2}})$,
$P_{\theta^{\ast}}$ in $\Gamma$ is replaced by $P_\theta$, and $R\ket
u=\ket{s^{-1}u}$ with $s^{-1}$ taken modulo $d$.

\subsection*{Privacy of unselected shares}
Acting with $\Uenc$ on an input $\ket\psi_{B_CE}$ (entangled with an
external system $E$ of any dimension) tensored with the key Bell pairs
$\ket\Phi_{S_jK_j}$ gives, by Eq.~\eqref{eq:weylsum},
$\Uenc\ket{\psi}\otimes\bigotimes_j\ket\Phi_{S_jK_j}
=\frac{1}{d^{2}}\sum_{p,q}\mu_{p,q}\ket{\chi_{p,q}}_{B_CE}\otimes\bigotimes_j\ket{\varphi_{p,q}}_{S_jK_j}$,
with $\ket{\chi_{p,q}}=(\Wpq\otimes\mathbb{I}_E)\ket\psi$ and
$\ket{\varphi_{p,q}}=(\Wpq\otimes\mathbb{I})\ket\Phi_{S_jK_j}$. By
Eq.~\eqref{eq:orthogonality}, $\braket{\varphi_{p',q'}}{\varphi_{p,q}}=
\delta_{pp'}\delta_{qq'}$ on each pair. Let $\mathcal C\subsetneq
\{1,\dots,n\}$ be any coalition of signal qudits that is not the whole
set, and let $j_0\notin\mathcal C$ be a spectator. Because
$\ket{\varphi_{p,q}}_{S_{j_0}K_{j_0}}$ is retained in the partial trace
and is orthogonal for distinct $(p,q)$, every cross term
$(p,q)\ne(p',q')$ is annihilated, and the reduced state of
$\mathcal C$ together with $E$ collapses to the incoherent mixture
\begin{equation}
\begin{aligned}
  \rho_{\mathcal C E}&=\sum_{p,q}\frac{|\mu_{p,q}|^{2}}{d^{4}}\,
  \Tr_{B_C}\!\big[\ket{\chi_{p,q}}\!\bra{\chi_{p,q}}\big]\\
  &\qquad\otimes\textstyle\bigotimes_{j\in\mathcal C}
  \Tr_{K_j}\!\big[\ket{\varphi_{p,q}}\!\bra{\varphi_{p,q}}\big]
  =\frac{\mathbb{I}^{\otimes|\mathcal C|}}{d^{|\mathcal C|}}\otimes\rho_E,
\end{aligned}
  \label{eq:coalition}
\end{equation}
using $\Tr_{B_C}[\ket{\chi_{p,q}}\!\bra{\chi_{p,q}}]=\rho_E$ (the Weyl
operator acts on $B_C$ alone and is removed by the partial trace, for
any $E$), $\Tr_{K_j}[\ket{\varphi_{p,q}}\!\bra{\varphi_{p,q}}]=\id/d$
for every $(p,q)$, and normalization
$\sum_{p,q}|\mu_{p,q}|^{2}=d^{4}$. No property of $\theta$ enters:
privacy of proper coalitions holds at every phase angle, and the
flat angle is required only for recovery. Every proper coalition
therefore holds a state that is maximally mixed \emph{and} uncorrelated
with $E$, for any $E$, every $d$ and every $n\ge2$. The argument
requires $n\ge2$: for $n=1$ the single share has no spectator, so
it is the all-shares coalition of Supplementary Section~\ref{si:coalition}, and it is private only
when $\gcd(1-\eta^{2},d)=1$. For $d=2,3,4,6$ it is correlated with $E$
(trace distance $1/2$ from $\id/d\otimes\rho_E$ at $d=2$), whereas for
$d=5,7$ it happens to be private. Recovery at $n=1$ remains exact.

\subsection*{Derivation of the closed-form decoder}
The decoder is obtained by inverting Eq.~\eqref{eq:weylsum} term by
term. Because $\Uenc$ is a sum over Weyl operators applied
\emph{identically} to $B_C$ and to every signal qudit, the post-encoding
state is a coherent superposition of $d^2$ branches labeled by $(p,q)$,
\begin{multline}
\Uenc\ket\Phi_{B_AB_C}\!\bigotimes_{j}\ket\Phi_{S_jK_j} \\
  =\frac{1}{d^{2}}\sum_{p,q}\mu_{p,q}\,
   \big(\Wpq\big)_{B_C}\ket\Phi_{B_AB_C}
   \bigotimes_{j=1}^{n}\big(\Wpq\otimes\mathbb I\big)\ket\Phi_{S_jK_j},
  \label{eq:branches}
\end{multline}
in which the \emph{same} label $(p,q)$ is shared by every register. The
decoder's task is therefore not to identify an abstract recovery map on
a $d^{2(n+1)}$-dimensional space, but to coherently transfer the shared
label from the $n$ signal--key pairs onto the two registers
$(S_B,K_B)$ and then undo it. Each step below is an exact operator
identity; no approximation or search enters.

\emph{Step 1 --- label extraction ($\CSUB(K_B\!\to\!S_B)$).}
On the pair $(S_B,K_B)$ the branch state is $(\Wpq\otimes\mathbb
I)\ket\Phi=d^{-1/2}\sum_k\omega^{qk}\ket{k+p}\ket{k}$. Subtracting the
key value from the signal value maps $\ket{k+p}\ket{k}\mapsto
\ket{p}\ket{k}$, factorizing the pair into
$\ket p_{S_B}\otimes F\ket q_{K_B}$: the shift label $p$ is now a
computational-basis value on $S_B$, and the phase label $q$ a Fourier
state on $K_B$.

\emph{Step 2 --- spectator restoration ($\CSUM(S_B\!\to\!K_j)$,
$j\ne B$).} Since $S_B$ holds $\ket p$, adding it into each spectator
key register maps $\ket{k+p}_{S_j}\ket{k}_{K_j}\mapsto
\ket{k+p}_{S_j}\ket{k+p}_{K_j}$, i.e.\ every unselected pair returns to
$\ket\Phi_{S_jK_j}$ up to the residual phase $\omega^{qk}$.

\emph{Step 3 --- phase collection ($\CSUM(K_j\!\to\!K_B)$ then
$\CSUB(S_B\!\to\!K_B)^{n}$).} Because $K_B$ holds a Fourier state,
$\CSUM(c\!\to\!K_B)$ acts on the control as $\Zdop^{-q}$ and leaves
$K_B$ untouched, while $\CSUB(c\!\to\!K_B)$ acts as $\Zdop^{+q}$; both
identities follow from $F^{-1}\Zdop F=\Xd$. Applying the first to each
of the $n-1$ spectator keys strips the phase $\omega^{-q(k+p)}$ off
each, canceling the residue of Step 2 and leaving all $n-1$ spectator
pairs in exactly $\ket\Phi_{S_jK_j}$; applying the second $n$ times to
$S_B$ removes the accumulated $\omega^{qp}$ bookkeeping. Steps 1--3 use
$3n-1$ two-qudit gates and together implement the exact map
\begin{equation}
\begin{split}
  \big(\Wpq &\otimes\mathbb I\big)^{\otimes n}\!\! \bigotimes_j\ket\Phi_{S_jK_j} \\
  &\longmapsto\;
  \omega^{pq}\,\ket p_{S_B}\otimes F\ket q_{K_B}
  \otimes\bigotimes_{j\ne B}\!\ket\Phi_{S_jK_j},
  \label{eq:extract}
\end{split}
\end{equation}
independently of $n$ --- the unselected sites are returned to their
original Bell pairs and never touched again.

\emph{Step 4 --- chirp cancellation ($\Lambda$ on $S_B$, $\Gamma$ on
$K_B$).} What remains is the branch weight $\mu_{p,q}=\hat g(p)\hat
g(q)=G^{2}\lambda^{-p^{2}}\lambda^{-q^{2}}$ from Eq.~\eqref{eq:chirp},
a product of two independent chirps. This is precisely why a
single-qudit diagonal gate suffices on each register:
$\Lambda=\mathrm{diag}(\lambda^{k^{2}})$ on $S_B$ removes
$\lambda^{-p^{2}}$, while $\Gamma=P_{\theta^{\ast}}^{\dagger}R\,
F^{\dagger}P_{\theta^{\ast}}^{\dagger}$ on $K_B$ obeys the identity
\begin{equation}
  \Gamma\,F\ket q=\frac{\overline G}{\sqrt d}\,\lambda^{q^{2}}\,F\ket{-q},
  \label{eq:gamma}
\end{equation}
where $\overline G$ is the complex conjugate of the Gauss sum $G$ of
Eq.~\eqref{eq:chirp}, removing $\lambda^{-q^{2}}$ and simultaneously
conjugating the phase label. Both gates are $q$- and $p$-independent, so they act identically
on every branch and the superposition is preserved. All $d^2$ branch
weights are now equal in modulus \emph{and} in phase, up to the global
constant $G^{2}\overline G/\sqrt d$; had the flatness of $|\hat g|$ been
only approximate, no such cancellation would exist, which is the sense
in which Eq.~\eqref{eq:chirp} is the structural fact the whole
construction rests on.

\emph{Step 5 --- recombination ($J$ on $S_B$, $\CSUM(K_B\!\to\!S_B)$).}
The state is now $\propto\sum_{p,q}\omega^{pq}(\Wpq)_{B_C}
\ket\Phi_{B_AB_C}\otimes\ket p_{S_B}\otimes F\ket{-q}_{K_B}$. Applying
$J\ket x=\ket{-x}$ to $S_B$ and then $\CSUM(K_B\!\to\!S_B)$ collapses
this double sum: the $p$-sum acts as a completeness relation transferring
$B_C$'s shift onto $S_B$, and the $q$-sum does the same for the phase.
The result is exactly
\begin{equation}
  \ket\Phi_{B_AS_B}\otimes\ket\Phi_{B_CK_B}
  \otimes\bigotimes_{j\ne B}\ket\Phi_{S_jK_j},
\end{equation}
which is Eq.~\eqref{eq:postdecode}: Alice's entanglement has been
transferred to the selected site with fidelity exactly one, and $B_C$
has been swapped into the key register $K_B$.

Counting gates, Steps 1--3 contribute $3n-1$ two-qudit gates, Step 4
two single-qudit gates, and Step 5 one single-qudit and one two-qudit
gate: $3n$ two-qudit and three single-qudit gates in total, matching
Eq.~\eqref{eq:decoder} and $O(n)$ in the number of sites; the
$(d,n)=(3,2)$ case is unrolled explicitly in Supplementary
Section~\ref{si:decoder}, and the full protocol's resource count is
tabulated in Supplementary Section~\ref{si:resources}. The
derivation uses no property of $B_A$, so it extends to an arbitrary
partner of $B_C$. For every $d\ge2$, $n\ge1$, flat angle $\theta$,
site $B$ and pure state $\ket\psi_{B_CE}$ with $E$ of any dimension,
\begin{equation}
\begin{split}
  \Udec^{(B)}\Uenc\,\ket\psi_{B_CE}&\bigotimes_j\ket\Phi_{S_jK_j}\\
  &=\ket\psi_{S_BE}\otimes\ket\Phi_{B_CK_B}\otimes
  \bigotimes_{j\ne B}\ket\Phi_{S_jK_j}.
\end{split}
  \label{eq:recovery}
\end{equation}
Steps 1--5 establish this for $\ket\psi=\ket\Phi_{B_CB'}$. A
general $\ket\psi_{B_CE}$ equals $\sqrt d\,(\id\otimes V)\ket\Phi_{B_CB'}$
for a linear map $V:B'\to E$, which commutes with the encoder and
decoder because they never act on $B'$
(Supplementary Section~\ref{si:cloning}). Eq.~\eqref{eq:postdecode}
also settles key reuse: the keys are left in Bell pairs with their own
signal qudits and retain no correlation with $B_A$, so a second site
applying the same decoder obtains entanglement fidelity $1/d^{2}$.

\subsection*{Deterministic non-local \texorpdfstring{$\CSUM^{k}$}{CSUM^k}}
We use the labels of the main text: data qudits $D_A$ (Alice) and $D_B$
(Bob), and a resource pair $\ket\Phi_{B_A\Sb}$. Alice applies
$\CSUM^{k_1}(D_A\!\to\!B_A)$ and Bob applies $\CSUB^{k_2}(\Sb\!\to\!D_B)$.
Alice then measures $B_A$ in the computational basis (outcome $m_A$), and
Bob applies $F$ to $\Sb$ and measures it in the computational basis
(outcome $m_B$). For all $k_1,k_2\in\Zd$ and every branch
$(m_A,m_B)$, the corrected branch operator on $D_AD_B$ is
\begin{equation}
\begin{split}
  M_{m_A,m_B}&:=\big(\Zdop^{k_1m_B}\!\otimes\Xd^{k_2m_A}\big)
  \big(\bra{m_A}_{B_A}\bra{m_B}_{\Sb}F_{\Sb}\big)\\
  &\qquad\times\CSUB^{k_2}_{\Sb\to D_B}\CSUM^{k_1}_{D_A\to B_A}
  \ket\Phi_{B_A\Sb}\\
  &=\frac{\omega^{m_Am_B}}{d}\,\CSUM^{k_1k_2}_{D_A\to D_B}.
\end{split}
  \label{eq:teleport}
\end{equation}
To see this, track a basis input $\ket c_{D_A}\ket t_{D_B}$; by
linearity this fixes the operator. After the two gates the state is
$d^{-1/2}\sum_j\ket c_{D_A}\ket{j+k_1c}_{B_A}\ket j_{\Sb}\ket{t-k_2j}_{D_B}$.
The outcome $m_A$ fixes $j=m_A-k_1c$, which leaves $D_B$ in
$\ket{t-k_2m_A+k_1k_2c}$ and $\Sb$ in $\ket{m_A-k_1c}$. Measuring $\Sb$
after $F$ gives each $m_B$ with probability $1/d$ and the phase
$\omega^{(m_A-k_1c)m_B}$. The corrections $\Xd^{k_2m_A}$ on $D_B$ and
$\Zdop^{k_1m_B}$ on $D_A$ remove the $m_A$ shift and the $c$-dependent
phase, and leave
\begin{equation}
  \ket c_{D_A}\ket{t+kc}_{D_B},\qquad k\equiv k_1k_2\pmod d,
\end{equation}
up to the global phase $\omega^{m_Am_B}$, in each of the $d^{2}$
equiprobable branches. The amplitude is $1/d$ for every $(c,t)$, which
establishes Eq.~\eqref{eq:teleport} and hence Eq.~\eqref{eq:csumk}. The protocol of
the main text is $k_1=k_2=1$, with corrections $\Xd^{m_A}$ and
$\Zdop^{m_B}$.

\emph{Altered classical messages.} Because the corrections are applied
after the measurements and act on $D_A$ and $D_B$ directly, an
alteration of the transcript composes with Eq.~\eqref{eq:teleport}
from the left. If Bob receives $m_A+\delta_A$ and Alice receives
$m_B+\delta_B$, branch $(m_A,m_B)$ implements
\begin{equation}
  \big(\Zdop^{k_1\delta_B}\otimes\Xd^{k_2\delta_A}\big)\,
  \CSUM^{k}(D_A\!\to\!D_B)
  \label{eq:tamper}
\end{equation}
exactly, still with probability $1/d^{2}$ and for every input. The
transcript is uniform and independent of the data, so a party
controlling the channel learns nothing from it, and the most it can do
is append a Weyl error of its choosing, determined by
$(\delta_A,\delta_B)$. That error is undone if the true transcript is
later authenticated. Authenticating the two dits
reduces the adversarial case to Eq.~\eqref{eq:main}.

Only the product $k_1k_2$ matters, so $\CSUM$ is obtained whenever
$k_1k_2\equiv1$: Bob may apply $\CSUM(\Sb\!\to\!D_B)$ ($k_2=-1$)
provided Alice applies $\CSUB(D_A\!\to\!B_A)$ ($k_1=-1$) and the
corrections are $\Xd^{-m_A}$ and $\Zdop^{-m_B}$. What fails is the
literal transcription of the qubit circuit, in which both parties apply
$\CSUM$ ($k_1=1$, $k_2=-1$): it teleports $\CSUB$, not $\CSUM$.
Applying the qubit correction $\Xd^{m_A}$ anyway leaves $D_B$ in $\ket{t-c+2m_A}$, displaced from the target
$\ket{t+c}$ by $2(m_A-c)$. This shift depends on the logical value $c$,
which is in general an unknown superposition, and no function of
$(m_A,m_B)$ can supply it without measuring $D_A$. At $d=2$ the
displacement is $0$ for every $c$, which is why the inversion is
invisible for qubits. For $d>2$ it vanishes only when
$m_A\equiv c\pmod{d/\nu(d)}$, i.e.\ on a fraction $\nu(d)/d$ of
(input, branch) pairs: $1/3$ at $d=3$, $1/2$ at $d=4$, $1/5$ at $d=5$
(Fig.~\ref{fig:obstruct}c, Supplementary Section~\ref{si:asymmetry}).
Which branches succeed depends on the unknown $c$, so the parties can
neither identify nor post-select them; the remedy is simply to fix the
signs so that $k_1k_2\equiv1$. The complete analysis, including
the resource count, is given in Supplementary
Section~\ref{si:teleport}.

\subsection*{Hiding the power of \texorpdfstring{$\CSUM$}{CSUM}}
With $k_2=1$, Bob's gate $\CSUB(\Sb\!\to\!D_B)$, his Fourier-basis
measurement and his correction $\Xd^{m_A}$ are the same for every $k$,
and the encoder, the decoder and the key register never involve $k$,
which enters only through Alice's local gate $\CSUM^{k}(D_A\!\to\!B_A)$
and her correction $\Zdop^{km_B}$. The public outcomes are uniform for
every $k$: in the derivation above $m_A=j+k_1c$ with $j$ uniform over
$\Zd$, and $m_B$ is uniform because $\Sb$ is in a computational-basis
state when it is measured in the Fourier basis. The joint distribution
of everything that Bob, the provider or an unselected site receives
from the protocol is therefore independent of $k$, and Alice may fix
$k$ at any time before her measurement. What the protocol cannot remove
is the action of the gate itself on Bob's data qudit. For a product
input, with Alice's computational-basis distribution
$p(c)=\bra c\rho_{D_A}\ket c$, the gate acts on $D_B$ as the channel
$\rho\mapsto\sum_c p(c)\,\Xd^{kc}\rho\,\Xd^{-kc}$, whatever Bob
prepares. Even with $D_B$ entangled with a reference he keeps, the
output is a mixture of the orthogonal states
$(\Xd^{kc}\otimes\id)\ket\Phi$, so the most Bob can extract is a single
sample of the product $kc$, drawn from Alice's input distribution.
Because that input is in general unknown to him, this sample does not
separate $k$ from $c$. If $p(c)$ is uniform, as for $D_A$ in any
Fourier-basis state or maximally entangled with an external system, and
$\gcd(k,d)=1$, the sample is uniform and $D_B$'s output is independent
of $k$ altogether. In general, repeated use can reveal at most which
subgroup $\gcd(k,d)\,\Zd$ the products $kc$ fall in. This is exactly
the leakage of an ideal local $\CSUM^{k}$, so the protocol hides $k$ as
well as any implementation of the gate could.

\subsection*{Derivation of the exact gate and timing of Alice's operation}
\emph{Purification.} Let $\ket{\rho}_{D_AD_BRR'}$ purify $\rho$ with an
additional reference $R'$. Every operation of the protocol acts as
the identity on $RR'$, so it suffices to show that each branch maps
$\ket\rho$ to $\frac{\omega^{m_Am_B}}{d}(\CSUM^{k}\otimes\id_{RR'})
\ket\rho$ times a fixed state of the remaining registers; tracing out
$R'$ then gives Eq.~\eqref{eq:main} for $\rho$.

\emph{Recovery.} Before Stage~3, the registers $D_A,D_B,R,R'$ have not
been acted on, and the joint state is
$\ket\rho\otimes\Udec^{(B)}\Uenc\ket\Phi_{B_AB_C}\bigotimes_j
\ket\Phi_{S_jK_j}$. Eq.~\eqref{eq:recovery} with $E=B_A$ turns the
second factor into
$\ket\Phi_{B_A\Sb}\otimes\ket\Phi_{B_CK_B}\otimes\bigotimes_{j\ne
B}\ket\Phi_{S_jK_j}$, a fixed state independent of $\rho$.

\emph{Teleportation.} Stage~3 acts only on $D_A,D_B,B_A,\Sb$. By
Eq.~\eqref{eq:teleport}, branch $(m_A,m_B)$ applies
$M_{m_A,m_B}\otimes\id_{RR'}=\frac{\omega^{m_Am_B}}{d}\CSUM^{k}
\otimes\id_{RR'}$ to $\ket\rho$ and leaves the other registers in
$\ket\Phi_{B_CK_B}\otimes\bigotimes_{j\ne B}\ket\Phi_{S_jK_j}$. The
branch probability is $\|M_{m_A,m_B}\ket\rho\|^{2}=1/d^{2}$ because
$\CSUM^{k}$ is unitary, and the normalized output is
$(\CSUM^{k}\otimes\id)\ket\rho$, up to a global phase, which is
Eq.~\eqref{eq:main}. Eq.~\eqref{eq:teleport} is an operator identity on $D_AD_B$,
so it applies unchanged to inputs entangled with $RR'$; nothing about
$R$ is used beyond the fact that the protocol does not act on it.

\emph{Timing.} Alice's gate $\CSUM^{k_1}(D_A\!\to\!B_A)$ acts only on
$(D_A,B_A)$, while the encoder and decoder act only on
$(B_C,\{S_j,K_j\})$. The two commute, so Alice may apply her gate in
Stage~1, before encoding, or in Stage~3, as in the main text, with the
same joint state; in the first case Eq.~\eqref{eq:recovery} is
applied with $E=B_AD_AD_BRR'$. The same disjointness carries the privacy
and single-use guarantees over to the composed protocol unchanged. It
also makes $m_A$ and $m_B$ safe to announce openly: each is a function
of a register the encoder never touched, or of $\Sb$ after decoding,
so neither carries information about any unselected share.

\subsection*{Practical consequences of the decoder}
\emph{Reuse of entanglement.} Only the provider's original pair is
consumed. After decoding, every unselected pair $(S_j,K_j)$ is back in
$\ket\Phi$, now shared between site $j$ and the recipient, and
$(B_C,K_B)$ is a fresh Bell pair between provider and recipient
(Eq.~\eqref{eq:postdecode}); all $n$ can be reused in later rounds. The
key is single-use, but the entanglement is not.

\emph{Replacing lost keys.} For $j\ne B$ the decoder uses $K_j$ only as
the partner of $S_j$ in a Bell pair, so a key lost in transit can be
replaced by the share $S_j$ of a site that has declined the gate,
provided it is still coherent. Because the two halves of a Bell pair
are interchangeable only up to a transpose, the decoder must be
adjusted (Supplementary Section~\ref{si:cloning}); with the adjustment
recovery is exact, and without it the fidelity falls to $1/d^{2}$.

\emph{Late binding of the data.} Alice's entangling gate commutes with
the encoder and decoder, which act on disjoint registers, so she may
apply it at the last moment, and $D_A$ need not exist when the resource
is distributed.

\subsection*{Noise model and figure of merit}
Each exposed register is acted on, between encoding and decoding, by
$\mathcal D_p(\rho)=(1-p)\rho+p\,\id\Tr[\rho]/d=(1-p+p/d^{2})\rho
+(p/d^{2})\sum_{(a,b)\ne(0,0)}W_{a,b}\rho W_{a,b}^{\dagger}$. The figure
of merit is the process fidelity of the completed gate: $D_A$ and $D_B$
start maximally entangled with reference qudits $R_A$, $R_B$ (unrelated
to the reference $R$ of Eq.~\eqref{eq:main}), and the
corrected output of each branch is compared, via the Uhlmann--Jozsa fidelity, with the Choi state
$\ket{C}=(\id_{R_AR_B}\otimes\CSUM)\ket\Phi_{R_AD_A}\ket\Phi_{D_BR_B}$,
\begin{equation}
  \mathcal F=\sum_{m_A,m_B}P(m_A,m_B)\,
  \bra{C}\rho^{(m_A,m_B)}_{R_AD_AD_BR_B}\ket{C}.
\end{equation}
Unlike a fidelity averaged over computational-basis inputs, this is
sensitive to phase errors, and a fully random gate scores $1/d^{2}$
rather than $1/d$. Because $\mathcal F$ is linear in each channel, noise
on $m$ registers gives $\mathcal F(p)=\sum_T(1-p)^{m-|T|}p^{|T|}\mathcal
F_T$, where $\mathcal F_T$ is the fidelity with the registers in $T$
fully depolarized, i.e.\ averaged uniformly over the $d^{2|T|}$ Weyl
errors. Each error is propagated as a pure state, so every curve in
Fig.~\ref{fig:noise} is exact. For a single register $\mathcal F_T$ is
$1/d^{2}$ for $\Sb$ and every $K_{j\ne B}$ (each non-trivial Weyl error
there reaches the data as a non-trivial Weyl error), $1$ for every
$S_{j\ne B}$, and $\gcd(n-\eta^{2},d)/d^{2}$ for $K_B$, with $\eta$ as
defined in the main text. The released key is special
because the decoder folds the other $n-1$ key labels into it: a Weyl
error $W_{a,b}$ on $K_B$ is harmless exactly when $a\equiv\eta b$ and
$(n-\eta^{2})b\equiv0$, which has $\gcd(n-\eta^{2},d)$ solutions. This
is the same arithmetic condition that governs the all-shares coalition.
It equals $1/d^{2}$ for every case in Fig.~\ref{fig:noise} ($n=2$,
$d\le4$), but not, for example, at $(d,n)=(2,3)$ or $(7,2)$.
Eq.~\eqref{eq:noiselinear} and its two-register analog follow from
these values. We verified them in the full protocol for $d=2$
($n\le5$), $d=3$ ($n\le4$), $d=4$ ($n\le3$) and $(d,n)=(5,2)$, and the
$K_B$ rule additionally in the cloning layer for $d=5,6,7$ with
$n\le4$ where feasible, including cases with $\gcd(n-\eta^{2},d)>1$.

\noindent\emph{Which registers matter.} After encoding, no operation of
the protocol acts on an unselected share $S_{j\ne B}$, so noise on it
commutes with everything that follows and disappears when it is traced
out; a provider need not protect the $n-1$ shares that will not be
redeemed. The keys $K_{j\ne B}$, by contrast, are decoder inputs, and
depolarizing any one of them is as damaging as depolarizing $\Sb$. If a lost key has been replaced by an unselected share (below),
that share joins the protected set. For two registers, of the $d^{4}$
joint Weyl errors on $(\Sb,K_B)$, $d^{2}$ cancel exactly in the decoder,
giving $\mathcal F(p)=1-(1-1/d^{2})[1-(1-p)^{2}]$ with floor $1/d^{2}$.
For general sets of noisy registers, multilinearity makes the
first-order term additive,
$1-\mathcal F\simeq p\sum_j(1-\mathcal F_{\{j\}})$, so exposing $\Sb$
and all $n$ keys gives
$1-\mathcal F\simeq\big[n(1-1/d^{2})+1-\gcd(n-\eta^{2},d)/d^{2}\big]p$
at small $p$, which is $(n+1)(1-1/d^{2})p$ when the gcd is $1$. The
pair floor of $1/d^{2}$ for $(\Sb,K_B)$ holds for every $(d,n)$ we
tested, including those with $\gcd(n-\eta^{2},d)>1$. Gates and
measurements are ideal throughout this analysis.

\subsection*{Independent numerical cross-check}
Every claim above is proved in closed form. As a check against
algebraic slips we also evaluated the construction numerically with
an exact state-vector implementation in NumPy, in which every gate,
measurement and correction acts directly on the qudit tensor.
The checks reproduce, to $<10^{-9}$ and typically $\sim10^{-15}$:
Eqs.~\eqref{eq:chirp}, \eqref{eq:weylsum}, \eqref{eq:extract}
and~\eqref{eq:gamma}; the flat-angle lattice, including a fine scan of
$\theta$; exact recovery for every site $B$ at $d\le9$ ($n=2$), $d\le6$
($n=3$) and $d\le4$ ($n=4$), and on every flat angle for $d\le9$, with
the post-decoding state Eq.~\eqref{eq:postdecode}; privacy of every
proper coalition ($d\le7$, $n\le4$); the full $(2n+4)$-qudit protocol,
including measurements and feed-forward, in every branch and for every
power $k$; the success fraction of the literal qubit transcription,
Eq.~\eqref{eq:uninvfrac}; the
access structure of Supplementary Section~\ref{si:access}, including
the symplectic map Eq.~\eqref{eq:symplectic}, the stabilizer group and
the counting formula Eq.~\eqref{eq:MIcount}; key replacement; the
$k$-independence of the public outcomes; the effect of altered
classical messages, Eq.~\eqref{eq:tamper}, for every $(k_1,k_2)$ and
alteration at $d\le5$; and the noise coefficients $\mathcal F_T$,
including the first-order scaling with the number of exposed keys.

\emph{Use of AI tools.} Large language models (Claude Opus, Anthropic)
were used to help write and refactor the numerical code and to draft the manuscript text. After using this tool, the author reviewed and edited the content as needed and take full responsibility for the content of the publication. Access to the code is described under Code availability.

\section*{Data availability}
No experimental datasets were generated or analyzed during this study.
The results reported here are analytic and follow from the
constructions given in Methods; the verification outputs quoted as
cross-checks (truth-table fidelities, coalition trace distances, and
the angle scan) are generated by the code described below\ and are
available on the same terms (Code availability).

\section*{Code availability}
The code used for the numerical cross-checks is not publicly
released because the protocol is the subject of a pending patent
application~\cite{Roy2026patent} (Competing interests). It is
registered as FERMILAB-CODE-2026-45-SQMS and will be made available to
editors and referees for peer review, and to other researchers under a
research license on request to the author or to Fermilab's Office of
Partnerships and Technology Transfer (OPTT@fnal.gov). It
consists of a NumPy module implementing the encoder
Eq.~\eqref{eq:encoder}, the closed-form decoder Eq.~\eqref{eq:decoder},
the full protocol and the noise model as exact state-vector operations,
and two Jupyter notebooks with their outputs. The first
checks every quantitative statement of this paper with an explicit
assertion. The second regenerates Figs.~\ref{fig:obstruct},
\ref{fig:noise} and~\ref{fig:privacy}, so every plotted quantity is
recomputed.

\begin{acknowledgments}
We acknowledge helpful feedback from Henry Lamm. This work was supported by the U.S. Department of Energy, Office of Science, National Quantum Information Science Research Centers, Superconducting Quantum Materials and Systems Center (SQMS), under Contract No. 89243024CSC000002. Fermilab is operated by Fermi Forward Discovery Group, LLC under Contract No. 89243024CSC000002 with the U.S. Department of Energy, Office of Science, Office of High Energy Physics.

\end{acknowledgments}

\section*{Author contributions}
T.R. conceived the project, developed the theory, carried out the
analytical derivations and numerical verification, and wrote the
manuscript.
\\

\section*{Competing interests}
T.R. is the inventor on a pending patent application covering the
protocol described in this work, U.S. Provisional Patent Application
No.~64/182,988~\cite{Roy2026patent}, filed by Fermi Forward Discovery
Group, LLC.

\bibliography{clone}

@article{kimble2008internet,
  title={The quantum internet},
  author={Kimble, H Jeff},
  journal={Nature},
  volume={453},
  number={7198},
  pages={1023--1030},
  year={2008},
  publisher={Nature Publishing Group UK London},
  url={https://doi.org/10.1038/nature07127},
  doi={10.1038/nature07127}
}

@article{wehner2018internet,
author = {Stephanie Wehner  and David Elkouss  and Ronald Hanson },
title = {Quantum internet: A vision for the road ahead},
journal = {Science},
volume = {362},
number = {6412},
pages = {eaam9288},
year = {2018},
doi = {10.1126/science.aam9288},
URL = {https://www.science.org/doi/abs/10.1126/science.aam9288}
}

@article{GottesmanChuang1999teleport,
  author = {Gottesman, Daniel and Chuang, Isaac L.},
  title = {Demonstrating the viability of universal quantum computation using teleportation and single-qubit operations},
  journal = {Nature},
  volume = {402},
  number = {6761},
  pages = {390--393},
  year = {1999},
  publisher = {Nature Publishing Group},
  doi = {10.1038/46503}
}

@article{Eisert2000,
  author = {Eisert, J. and Jacobs, K. and Papadopoulos, P. and Plenio, M. B.},
  title = {Optimal local implementation of nonlocal quantum gates},
  journal = {Phys. Rev. A},
  volume = {62},
  issue = {5},
  pages = {052317},
  year = {2000},
  publisher = {American Physical Society},
  doi = {10.1103/PhysRevA.62.052317}
}

@article{Chou2018teleport,
  author = {Chou, Kevin S. and Blumoff, Jacob Z. and Wang, Christopher S. and Reinhold, Philip C. and Axline, Christopher J. and Gao, Yvonne Y. and Frunzio, L. and Devoret, M. H. and Jiang, Liang and Schoelkopf, R. J.},
  title = {Deterministic teleportation of a quantum gate between two logical qubits},
  journal = {Nature},
  volume = {561},
  number = {7723},
  pages = {368--373},
  year = {2018},
  publisher = {Nature Publishing Group},
  doi = {10.1038/s41586-018-0470-y}
}

@article{YamaguchiKempf2026cloning,
  title = {Encrypted Qubits Can Be Cloned},
  author = {Yamaguchi, Koji and Kempf, Achim},
  journal = {Phys. Rev. Lett.},
  volume = {136},
  issue = {1},
  pages = {010801},
  numpages = {7},
  year = {2026},
  month = {Jan},
  publisher = {American Physical Society},
  doi = {10.1103/y4y1-1ll6},
  url = {https://link.aps.org/doi/10.1103/y4y1-1ll6}
}

@article{Hillery1999sharing,
  author = {Hillery, Mark and Bu\v{z}ek, Vladim\'{\i}r and Berthiaume, André},
  title = {Quantum secret sharing},
  journal = {Phys. Rev. A},
  volume = {59},
  issue = {3},
  pages = {1829--1834},
  year = {1999},
  publisher = {American Physical Society},
  doi = {10.1103/PhysRevA.59.1829}
}

@article{Cleve1999share,
  author = {Cleve, Richard and Gottesman, Daniel and Lo, Hoi-Kwong},
  title = {How to share a quantum secret},
  journal = {Phys. Rev. Lett.},
  volume = {83},
  issue = {3},
  pages = {648--651},
  year = {1999},
  publisher = {American Physical Society},
  doi = {10.1103/PhysRevLett.83.648}
}

@inproceedings{Ambainis2000private,
  author = {Ambainis, Andris and Mosca, Michele and Tapp, Alain and de Wolf, Ronald},
  booktitle = {Proceedings 41st Annual Symposium on Foundations of Computer Science},
  title = {Private quantum channels},
  year = {2000},
  pages = {547--553},
  organization = {IEEE},
  doi = {10.1109/SFCS.2000.892131}
}

@incollection{Broadbent2013,
  author = {Broadbent, Anne and Gutoski, Gus and Stebila, Douglas},
  booktitle = {Advances in Cryptology -- CRYPTO 2013},
  title = {Quantum One-Time Programs},
  year = {2013},
  publisher = {Springer Berlin Heidelberg},
  address = {Berlin, Heidelberg},
  pages = {344--360},
  isbn = {978-3-642-40041-8}
}

@article{Gottesman1999,
  author = {Gottesman, Daniel},
  title = {Fault-tolerant quantum computation with higher-dimensional systems},
  journal = {Chaos, Solitons \& Fractals},
  volume = {10},
  number = {10},
  pages = {1749--1758},
  year = {1999},
  publisher = {Elsevier},
  doi = {10.1016/S0960-0779(99)00098-6}
}

@article{Wootters1982no_cloning_qubit,
  author = {Wootters, W. K. and Zurek, W. H.},
  title = {A single quantum cannot be cloned},
  journal = {Nature},
  volume = {299},
  pages = {802--803},
  year = {1982},
  doi = {10.1038/299802a0}
}

@article{Chiribella2008no_cloning_gate,
  author = {Chiribella, G. and D'Ariano, G. M. and Perinotti, P.},
  title = {Optimal Cloning of Unitary Transformation},
  journal = {Phys. Rev. Lett.},
  volume = {101},
  pages = {180504},
  year = {2008},
  doi = {10.1103/PhysRevLett.101.180504}
}

@article{Ceara2026qudit,
  author = {Ceară, Filip-Ioan},
  title = {Cloning encrypted quantum states in arbitrary dimensions},
  journal = {Phys. Rev. A},
  volume = {114},
  pages = {022450},
  year = {2026},
  doi = {10.1103/b32z-r5p7}
}

@misc{LimLo2026AME,
  author = {Lim, Zheng Liang and Lo, Hoi-Kwong},
  title = {Encrypted Cloning, Absolute Maximal Entanglement and Quantum Secret Sharing},
  year = {2026},
  eprint = {2605.26866},
  archivePrefix = {arXiv},
  primaryClass = {quant-ph},
}

@article{Shamir1979share,
  author = {Shamir, Adi},
  title = {How to share a secret},
  journal = {Commun. ACM},
  volume = {22},
  number = {11},
  pages = {612--613},
  year = {1979},
  doi = {10.1145/359168.359176}
}

@article{NielsenChuang1997programmable,
  author = {Nielsen, Michael A. and Chuang, Isaac L.},
  title = {Programmable quantum gate arrays},
  journal = {Phys. Rev. Lett.},
  volume = {79},
  pages = {321--324},
  year = {1997},
  doi = {10.1103/PhysRevLett.79.321}
}

@article{Huelga2001remote,
  author = {Huelga, S. F. and Vaccaro, J. A. and Chefles, A. and Plenio, M. B.},
  title = {Quantum remote control: Teleportation of unitary operations},
  journal = {Phys. Rev. A},
  volume = {63},
  pages = {042303},
  year = {2001},
  doi = {10.1103/PhysRevA.63.042303}
}

@article{Gheorghiu2014standard,
  author = {Gheorghiu, Vlad},
  title = {Standard form of qudit stabilizer groups},
  journal = {Phys. Lett. A},
  volume = {378},
  number = {5--6},
  pages = {505--509},
  year = {2014},
  doi = {10.1016/j.physleta.2013.12.009}
}

@article{Wang2007controlled,
  author = {Wang, An Min},
  title = {Combined and controlled remote implementations of partially unknown quantum operations of multiqubits using {G}reenberger-{H}orne-{Z}eilinger states},
  journal = {Phys. Rev. A},
  volume = {75},
  pages = {062323},
  year = {2007},
  doi = {10.1103/PhysRevA.75.062323}
}

@article{SaiToh2009threshold,
  author = {SaiToh, Akira and Rahimi, Robabeh and Nakahara, Mikio},
  title = {Economical $(k,m)$-threshold controlled quantum teleportation},
  journal = {Phys. Rev. A},
  volume = {79},
  pages = {062313},
  year = {2009},
  doi = {10.1103/PhysRevA.79.062313}
}

@misc{Gianini2026leak,
  author = {Gianini, Gabriele and Hasan, Omar and Mio, Corrado and Cimato, Stelvio and Damiani, Ernesto},
  title = {Encrypted clones can leak: Classification of informative subsets in quantum encrypted cloning},
  year = {2026},
  eprint = {2604.10155},
  archivePrefix = {arXiv},
  primaryClass = {quant-ph}
}

@misc{Gianini2026full,
  author = {Gianini, Gabriele and Cimato, Stelvio and Lin, Jianyi and Hasan, Omar and Damiani, Ernesto},
  title = {Full characterization of informative subsets in quantum encrypted cloning},
  year = {2026},
  eprint = {2605.27421},
  archivePrefix = {arXiv},
  primaryClass = {quant-ph}
}

@misc{Bai2026classification,
  author = {Bai, Chen-Ming and Zhou, Xin-Liang and Luo, Yu},
  title = {Classification of informative subsets in quantum encrypted cloning on qudits},
  year = {2026},
  eprint = {2605.11642},
  archivePrefix = {arXiv},
  primaryClass = {quant-ph}
}

@article{Reznik2002remote,
  author = {Reznik, Benni and Aharonov, Yakir and Groisman, Berry},
  title = {Remote operations and interactions for systems of arbitrary-dimensional {H}ilbert space},
  journal = {Phys. Rev. A},
  volume = {65},
  pages = {032312},
  year = {2002},
  doi = {10.1103/PhysRevA.65.032312}
}

@article{Karlsson1999sharing,
  author = {Karlsson, Anders and Koashi, Masato and Imoto, Nobuyuki},
  title = {Quantum entanglement for secret sharing and secret splitting},
  journal = {Phys. Rev. A},
  volume = {59},
  pages = {162--168},
  year = {1999},
  doi = {10.1103/PhysRevA.59.162}
}

@article{Keet2010qudit,
  author = {Keet, Adrian and Fortescue, Ben and Markham, Damian and Sanders, Barry C.},
  title = {Quantum secret sharing with qudit graph states},
  journal = {Phys. Rev. A},
  volume = {82},
  pages = {062315},
  year = {2010},
  doi = {10.1103/PhysRevA.82.062315}
}

@article{Helwig2012AME,
  author = {Helwig, Wolfram and Cui, Wei and Latorre, Jos\'e Ignacio and Riera, Arnau and Lo, Hoi-Kwong},
  title = {Absolute maximal entanglement and quantum secret sharing},
  journal = {Phys. Rev. A},
  volume = {86},
  pages = {052335},
  year = {2012},
  doi = {10.1103/PhysRevA.86.052335}
}

@article{Ogawa2005ramp,
  author = {Ogawa, Tomohiro and Sasaki, Akira and Iwamoto, Mitsugu and Yamamoto, Hirosuke},
  title = {Quantum secret sharing schemes and reversibility of quantum operations},
  journal = {Phys. Rev. A},
  volume = {72},
  pages = {032318},
  year = {2005},
  doi = {10.1103/PhysRevA.72.032318}
}

@article{Bell2014graph,
  author = {Bell, B. A. and Markham, D. and Herrera-Mart\'{\i}, D. A. and Marin, A. and Wadsworth, W. J. and Rarity, J. G. and Tame, M. S.},
  title = {Experimental demonstration of graph-state quantum secret sharing},
  journal = {Nat. Commun.},
  volume = {5},
  pages = {5480},
  year = {2014},
  doi = {10.1038/ncomms6480}
}

@article{Pinnell2020elevendim,
  author = {Pinnell, Jonathan and Nape, Isaac and de Oliveira, Micha{\"e}l and TabeBordbar, Najmeh and Forbes, Andrew},
  title = {Experimental demonstration of 11-dimensional 10-party quantum secret sharing},
  journal = {Laser Photonics Rev.},
  volume = {14},
  pages = {2000012},
  year = {2020},
  doi = {10.1002/lpor.202000012}
}

@inproceedings{Aaronson2009copy,
  author = {Aaronson, Scott},
  title = {Quantum copy-protection and quantum money},
  booktitle = {Proceedings of the 24th Annual IEEE Conference on Computational Complexity},
  pages = {229--242},
  year = {2009},
  doi = {10.1109/CCC.2009.42}
}

@inproceedings{BroadbentLord2020uncloneable,
  author = {Broadbent, Anne and Lord, S\'ebastien},
  title = {Uncloneable quantum encryption via oracles},
  booktitle = {15th Conference on the Theory of Quantum Computation, Communication and Cryptography (TQC 2020)},
  series = {LIPIcs},
  volume = {158},
  pages = {4:1--4:22},
  year = {2020},
  doi = {10.4230/LIPIcs.TQC.2020.4}
}

@inproceedings{Christandl2005anonymous,
  author = {Christandl, Matthias and Wehner, Stephanie},
  title = {Quantum anonymous transmissions},
  booktitle = {Advances in Cryptology -- ASIACRYPT 2005},
  series = {Lecture Notes in Computer Science},
  volume = {3788},
  pages = {217--235},
  year = {2005},
  publisher = {Springer},
  doi = {10.1007/11593447_12}
}

@inproceedings{Crepeau2002multiparty,
  author = {Cr\'epeau, Claude and Gottesman, Daniel and Smith, Adam},
  title = {Secure multi-party quantum computation},
  booktitle = {Proceedings of the 34th Annual ACM Symposium on Theory of Computing (STOC)},
  pages = {643--652},
  year = {2002},
  doi = {10.1145/509907.510000}
}

@inproceedings{Broadbent2009blind,
  author = {Broadbent, Anne and Fitzsimons, Joseph and Kashefi, Elham},
  title = {Universal blind quantum computation},
  booktitle = {50th Annual IEEE Symposium on Foundations of Computer Science (FOCS)},
  pages = {517--526},
  year = {2009},
  doi = {10.1109/FOCS.2009.36}
}

@article{Main2025distributed,
  author = {Main, D. and Drmota, P. and Nadlinger, D. P. and Ainley, E. M. and Agrawal, A. and Nichol, B. C. and Srinivas, R. and Araneda, G. and Lucas, D. M.},
  title = {Distributed quantum computing across an optical network link},
  journal = {Nature},
  volume = {638},
  pages = {383--388},
  year = {2025},
  doi = {10.1038/s41586-024-08404-x}
}

@article{Wan2019teleport,
  author = {Wan, Yong and Kienzler, Daniel and Erickson, Stephen D. and Mayer, Karl H. and Tan, Ting Rei and Wu, Jenny J. and Vasconcelos, Hilma M. and Glancy, Scott and Knill, Emanuel and Wineland, David J. and Wilson, Andrew C. and Leibfried, Dietrich},
  title = {Quantum gate teleportation between separated qubits in a trapped-ion processor},
  journal = {Science},
  volume = {364},
  pages = {875--878},
  year = {2019},
  doi = {10.1126/science.aaw9415}
}

@article{Ringbauer2022qudit,
  author = {Ringbauer, Martin and Meth, Michael and Postler, Lukas and Stricker, Roman and Blatt, Rainer and Schindler, Philipp and Monz, Thomas},
  title = {A universal qudit quantum processor with trapped ions},
  journal = {Nat. Phys.},
  volume = {18},
  pages = {1053--1057},
  year = {2022},
  doi = {10.1038/s41567-022-01658-0}
}

@article{Appleby2005clifford,
  author = {Appleby, D. M.},
  title = {Symmetric informationally complete--positive operator valued measures and the extended {C}lifford group},
  journal = {J. Math. Phys.},
  volume = {46},
  pages = {052107},
  year = {2005},
  doi = {10.1063/1.1896384}
}

@article{Cerf2002dlevel,
  author = {Cerf, Nicolas J. and Bourennane, Mohamed and Karlsson, Anders and Gisin, Nicolas},
  title = {Security of quantum key distribution using d-level systems},
  journal = {Phys. Rev. Lett.},
  volume = {88},
  pages = {127902},
  year = {2002},
  doi = {10.1103/PhysRevLett.88.127902}
}

@misc{huang2026sideband,
      title={Universal Jaynes-Cummings Control of an Oscillator}, 
      author={Jordan Huang and Ethan Kasaba and Thomas J. DiNapoli and Tanay Roy and Srivatsan Chakram},
      year={2026},
      eprint={2605.18658},
      archivePrefix={arXiv},
      primaryClass={quant-ph}, 
}

@misc{kim2026ultracoherent,
      title={Ultracoherent superconducting cavity-based multiqudit platform with error-resilient control}, 
      author={Taeyoon Kim and Tanay Roy and Xinyuan You and Andy C. Y. Li and Henry Lamm and Oleg Pronitchev and Mustafa Bal and Sabrina Garattoni and Francesco Crisa and Daniel Bafia and Doga Kurkcuoglu and Roman Pilipenko and Paul Heidler and Nicholas Bornman and David van Zanten and Silvia Zorzetti and Roni Harnik and Akshay Murthy and Andrei Lunin and Sergey Belomestnykh and Shaojiang Zhu and Changqing Wang and Andre Vallieres and Ziwen Huang and Jens Koch and Anna Grassellino and Srivatsan Chakram and Alexander Romanenko and Yao Lu},
      year={2025},
      eprint={2506.03286},
      archivePrefix={arXiv},
      primaryClass={quant-ph}, 
}

@inproceedings{Barnum2002authentication,
  author = {Barnum, Howard and Cr{\'e}peau, Claude and Gottesman, Daniel and Smith, Adam and Tapp, Alain},
  booktitle = {Proceedings 43rd Annual IEEE Symposium on Foundations of Computer Science},
  title = {Authentication of quantum messages},
  year = {2002},
  pages = {449--458},
  organization = {IEEE},
  doi = {10.1109/SFCS.2002.1181969}
}

@article{WegmanCarter1981,
  author = {Wegman, Mark N. and Carter, J. Lawrence},
  title = {New hash functions and their use in authentication and set equality},
  journal = {Journal of Computer and System Sciences},
  volume = {22},
  number = {3},
  pages = {265--279},
  year = {1981},
  doi = {10.1016/0022-0000(81)90033-7}
}

@article{Zhang2011shared,
doi = {10.1088/0953-4075/44/16/165508},
url = {https://doi.org/10.1088/0953-4075/44/16/165508},
year = {2011},
month = {jul},
publisher = {},
volume = {44},
number = {16},
pages = {165508},
author = {Zhang, Zhan-jun and Cheung, Chi-Yee},
title = {Shared quantum remote control: quantum operation sharing},
journal = {Journal of Physics B: Atomic, Molecular and Optical Physics}
}

@article{bornman2025benchmarking,
  title={Benchmarking the performance of a high-Q cavity qudit using random unitaries},
  author={Bornman, Nicholas and Roy, Tanay and Job, Joshua A and Anand, Namit and Perdue, Gabriel N and Zorzetti, Silvia and Alam, M Sohaib},
  journal={Quantum Science and Technology},
  volume={10},
  number={2},
  pages={025062},
  year={2025},
  publisher={IOP Publishing},
  url={https://doi.org/10.1088/2058-9565/adc47c},
  doi={10.1088/2058-9565/adc47c}
}

@misc{Roy2026patent,
  author = {Roy, Tanay},
  title = {Systems and methods for deferred-recipient teleportation of a two-qudit gate},
  howpublished = {U.S. Provisional Patent Application No. 64/182,988},
  year = {2026}
}


\renewcommand{\thetable}{S\arabic{table}}
\renewcommand{\theHtable}{S\arabic{table}}
\setcounter{table}{0}
\renewcommand{\thefigure}{S\arabic{figure}}
\renewcommand{\theHfigure}{S\arabic{figure}}
\setcounter{figure}{0}
\renewcommand{\theequation}{S\arabic{equation}}
\renewcommand{\theHequation}{S\arabic{equation}}
\setcounter{equation}{0}

\renewcommand{\thesubsection}{S\arabic{subsection}}
\setcounter{subsection}{0} 

\clearpage
\newpage

\textbf{Supplementary Information}

\subsection{Extended qudit gate identities}
\label{si:identities}
Beyond the commutation relation Eq.~\eqref{eq:comm} (Methods) and the Weyl
orthogonality relation Eq.~\eqref{eq:orthogonality} (Methods) used throughout the
main text, the following identities are used implicitly at various
points and are collected here for reference. With $\Wpq=\Xd^p\Zdop^q$,
the Weyl operators compose as
\begin{equation}
  W_{p,q}\,W_{p',q'}=\omega^{qp'}\,W_{p+p',q+q'},
  \label{eq:weylcompose}
\end{equation}
obtained by commuting $\Zdop^q$ through $\Xd^{p'}$ using
Eq.~\eqref{eq:comm} (Methods). The Fourier transform exchanges the two
generators in both directions, $F^{\dagger}\Zdop F=\Xd$ and
$F\,\Xd\,F^{\dagger}=\Zdop$, consistent with $F^4 = \mathbb{I}$. For any
operator $M$ on a single qudit and the maximally entangled state
$\ket\Phi=d^{-1/2}\sum_j\ket{jj}$, the transpose trick
$(M\otimes\mathbb{I})\ket\Phi=(\mathbb{I}\otimes M^{T})\ket\Phi$ holds in the
computational basis, together with $\bra\Phi(M\otimes\mathbb{I})\ket\Phi
=\Tr[M]/d$; both are used repeatedly when moving an operator from one
half of a Bell pair to the other, as in the extension of exact recovery
to arbitrary inputs (Methods, Eq.~\eqref{eq:recovery}, and
Section~\ref{si:cloning}). Finally,
$\CSUB:=\CSUM^{-1}=\CSUM^\dagger$ since $\CSUM$ is a
computational-basis permutation matrix (real and orthogonal), so
inverting it costs nothing beyond relabelling the gate.

\subsection{Why the qubit encoder does not generalize algebraically}
\label{si:obstruction}
For $d=2$, $\Xd^2=\Zdop^2=\id$, so the Weyl group collapses to just
four distinct operators, $\{\id,X,Z,XZ\}$ (with $XZ\propto Y$ up to
phase), and every one of them is again a single-qubit Pauli string.
This is what lets the qubit encoder be written compactly as a product
of two exponentials of Pauli strings: any diagonal phase built from
squares of sums of $Z$'s expands, using $Z^2=\id$, into a sum of purely
pairwise $Z_iZ_j$ terms with fixed coefficients, which exponentiates
directly into a product of two-qubit phase gates. For $d>2$, $\Xd^d=
\Zdop^d=\id$ but no smaller power vanishes, so $\{W_{p,q}\}_{p,q\in\Zd}$
is a full set of $d^2$ \emph{distinct} operators rather than a group of
order $4$: for $d=3$, for instance, direct computation confirms
$\Xd\Zdop$ is neither proportional to $\id,\Xd,\Zdop,\Xd^2,\Zdop^2$ nor
to any other single generator, and the analogous expansion of
$e^{i\theta T^2}$ (with $T$ a sum of $d$-level registers) does not
collapse to pairwise terms with $d$-independent coefficients, because
$Z_d^2\ne\id$ leaves genuinely higher powers $Z_d^{2k}$ in the
expansion for $k=1,\dots,\lfloor(d-1)/2\rfloor$. This is why the
construction here abandons the two-exponential route entirely in favor
of the explicit $\CSUM$-ladder circuit of Eq.~\eqref{eq:encoder}, whose
cost is $O(n)$ for every $d$ by direct construction rather than by
algebraic collapse.

\subsection{Full unrolled protocol circuit}
\label{si:fullcircuit}
\begin{figure*}[t]
\centering
\figbox{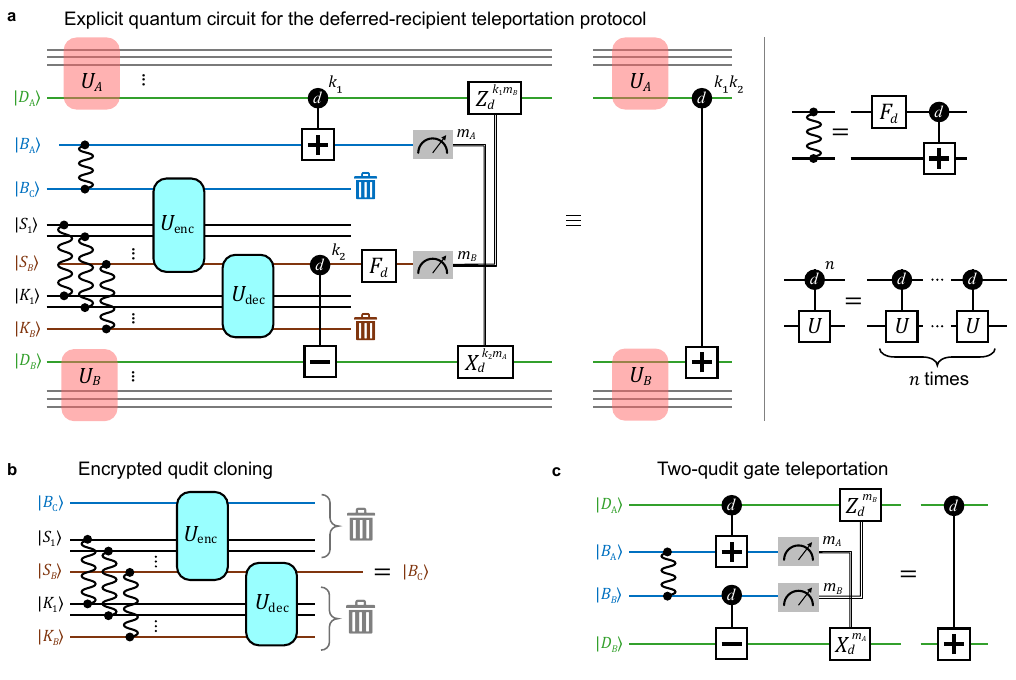}{\textwidth}
\caption{\textbf{The protocol unrolled into elementary qudit gates.}
\textbf{a}, The complete deferred-recipient construction on all $2n+4$
registers. Vertical wavy connectors joining two wires denote a Bell pair shared
between those two registers, prepared by the $F_d$-then-$\CSUM$ gadget
shown at the right: the provider's pair $(B_A,B_C)$, and one pair
$(S_j,K_j)$ for each candidate site. The data qudits $D_A$ and
$D_B$ enter through arbitrary unitaries $U_A$, $U_B$ (red) that entangle
them with external wires; the protocol never references those wires,
which is why the gate is exact on arbitrarily entangled inputs. The
provider encodes $B_C$ into the shares with $\Uenc$; site $B$ redeems
the released key and applies $\Udec$; $B_C$, the unselected shares and
the retained keys are then discarded (bins). A computational-basis
measurement on $B_A$ and a Fourier-basis measurement on $S_B$ return
dits $m_A$, $m_B$, which drive $Z_d^{m_B}$ on $D_A$ and $X_d^{m_A}$ on
$D_B$. The result equals a bare $\CSUM(D_A\!\to\!D_B)$ acting through
the same $U_A$, $U_B$, in every one of the $d^{2}$ branches.
\textbf{b}, The cloning layer alone: $\Udec$ inverts $\Uenc$ on the
redeeming site's share and key, returning $\ket{B_C}$ exactly.
\textbf{c}, The teleportation layer alone, with the resource pair
supplied directly. Bob's entangling gate is $\CSUB$; running it as
$\CSUM$ while keeping Alice's gate and the corrections unchanged
succeeds on only a $\nu(d)/d$ fraction of entries
(Section~\ref{si:asymmetry}).}
\label{fig:full_circuit}
\end{figure*}

Figure~\ref{fig:full_circuit}\textbf{a} unrolls the entire protocol into
elementary qudit gates on all $2n+4$ registers, with the encoder and
decoder shown as the blocks $\Uenc$ and $\Udec$ whose interiors are
given explicitly in Fig.~\ref{fig:circuits}. The data qudits $D_A$ and
$D_B$ are acted on by arbitrary unitaries $U_A$ and $U_B$ that entangle
them with external systems, drawn as extra wires; these unitaries are
never referenced by the protocol, which is what makes the gate exact on
arbitrary entangled inputs rather than only on product inputs. The
unselected shares $S_{j\ne B}$, the retained keys $K_{j\ne B}$, and the
provider's copy $B_C$ are discarded, marked by the bins: no part of the
protocol conditions on them, and by
Eq.~\eqref{eq:coalition} they carry nothing to condition on. The
equivalence sign asserts that the whole left-hand side reduces to a
single $\CSUM$ between $D_A$ and $D_B$ acting through the same $U_A$,
$U_B$.

Panels \textbf{b} and \textbf{c} isolate the two restrictions discussed
in the main text, and are analyzed in Sections~\ref{si:cloning} and~\ref{si:teleport}. Panel
\textbf{b} keeps only the cloning layer: $\Uenc$ followed by $\Udec$
returns $\ket{B_C}$ to the redeeming site exactly, which is the qudit
encrypted-cloning statement. Panel \textbf{c} keeps only the
teleportation layer, with the resource pair $(B_A,B_B)$ supplied
directly rather than recovered from a share. Bob's entangling gate is
$\CSUB(B_B\!\to\!D_B)$ ($k_1=k_2=1$ in Eq.~\eqref{eq:csumk}); the
role of this sign is discussed in Section~\ref{si:asymmetry}.

\subsection{Decoder: worked example}
\label{si:decoder}
The decoder derivation in Methods is complete as stated; this section
gives the $(d,n)=(3,2)$ case explicitly as a check on the index
conventions. At $d=3$ (odd), $\nu(3)=\gcd(2,3)=1$ and
$\nu_{\rm inv}(3)=2/\nu(3)=2$; $\theta^{\ast}=2\pi/3$,
$\nu_{\rm inv}^{-1}=2\bmod 3$, and $\lambda=\omega^{4^{-1}}=\omega$ since
$4\equiv1\bmod 3$. Hence
$\Lambda=\mathrm{diag}(1,\omega,\omega)$ (from $\lambda^{k^{2}}$ with
$k^{2}=0,1,1$), $R=\mathrm{diag\text{-}perm}\ket u\mapsto\ket{2u}$,
i.e.\ the transposition $(1\,2)$, and
$\Gamma=P^{\dagger}RF^{\dagger}P^{\dagger}$ with
$P=\mathrm{diag}(1,\omega,\omega)$. Taking $B=1$, Eq.~\eqref{eq:decoder}
unrolls to the six two-qudit gates
\begin{equation}
\begin{aligned}
  &\CSUB(K_1\!\to\!S_1),\quad \CSUM(S_1\!\to\!K_2),\quad
   \CSUM(K_2\!\to\!K_1),\\
  &\CSUB(S_1\!\to\!K_1)^{2},\quad \CSUM(K_1\!\to\!S_1),
\end{aligned}
\end{equation}
interleaved with $\Lambda_{S_1}$, $\Gamma_{K_1}$ and $J_{S_1}$ as in
Eq.~\eqref{eq:decoder}, matching $3n=6$ two-qudit and three
single-qudit gates. Applying this to Eq.~\eqref{eq:branches} reproduces
Eq.~\eqref{eq:postdecode} exactly. Gate-level circuit diagrams in the
style of Fig.~\ref{fig:circuits}b for $(d,n)=(2,2)$ and $(3,2)$ follow
mechanically from the same unrolling.

\subsection{Resource count vs.\ \texorpdfstring{$n$}{n} and \texorpdfstring{$d$}{d}}
\label{si:resources}
The complete protocol (Fig.~\ref{fig:full_circuit}a) uses $2n+4$ qudits: $D_A,D_B,B_A$, the encoded
qudit (relabelled $S_1,\dots,S_n$ after encoding), and the $n$ key
qudits $K_1,\dots,K_n$. Table~\ref{tab:resources} accounts for gate and
measurement cost as a function of $n$. Every entry is independent of
$d$, including the decoder: with Eq.~\eqref{eq:decoder} in hand its
cost is a gate count, $O(n)$, rather than the $d^{2(n+1)}$
Hilbert-space dimension on which a generic recovery map would have to
be synthesised. The whole protocol is therefore linear in the number of
sites and constant in the qudit dimension, so the only quantity that
grows with $d$ is the cost of implementing a single elementary qudit
gate on hardware.
\begin{table}[h]
\centering
\caption{Resource count vs.\ $n$, independent of $d$ throughout.
``CSUM-type'' counts elementary $\CSUM$ or $\CSUB$ gates;
``Fourier layers'' counts applications of $F^{\otimes(n+1)}$ or
$F^{-\otimes(n+1)}$.}
\label{tab:resources}
\begin{tabular}{@{}lc@{}}
\toprule
Resource & Count\\
\midrule
Total qudits & $2n+4$\\
Encoder: CSUM-type gates & $4n$\\
Encoder: phase gates & $2$\\
Encoder: Fourier layers & $2$\\
Decoder: CSUM-type gates & $3n$\\
Decoder: single-qudit gates & $3$\\
Entangling gates (teleportation) & $2$\\
Projective measurements & $2$\\
Classical dits exchanged & $2$\\
\bottomrule
\end{tabular}
\end{table}

\subsection{The flat-angle lattice}
\label{si:angle}
Methods shows that at $\theta^{\ast}(d)$ the Gauss sum is the pure
chirp $\hat g(p)=G\lambda^{-p^{2}}$, so $|\hat g(p)|=\sqrt d$ is an
identity rather than a coincidence of small $d$. Here we record which
\emph{other} angles share that property, since $\theta^{\ast}$ is the
smallest flat angle but not the only one.

Write $\theta=\pi s/d$ with $s\in\mathbb Z$, so that
$g(k)=e^{i\pi sk^{2}/d}$ and $\hat g(p)=\sum_k e^{i\pi(sk^{2}-2pk)/d}$.
Re-indexing $k\mapsto k+d$ leaves $g$ invariant iff $e^{i\pi sd}=1$,
i.e.\ iff $sd$ is even. Given that, and $\gcd(s,d)=1$, the square can be
completed. For even $d$, $s$ is odd and hence invertible modulo $2d$,
with inverse $\bar s$, and $sk^{2}-2pk\equiv s(k-\bar sp)^{2}-\bar sp^{2}
\pmod{2d}$. For odd $d$, $s$ is even and $2s$ is invertible modulo $d$,
and the same step runs in $\mathbb Z_d$ with $\omega$. Either way
$\hat g(p)=G_s\lambda_s^{-p^{2}}$ with $G_s=\sum_k
e^{i\pi sk^{2}/d}$ and $\lambda_s$ as in Eq.~\eqref{eq:lambdas}. Note
that $\lambda_s\neq e^{i\pi s/d}$ in general: at $\theta^{\ast}$ for odd
$d$, $\lambda=\omega^{4^{-1}}$, not $\omega$. The classical evaluation of
the quadratic Gauss sum gives $|G_s|=\sqrt d$ exactly when
$\gcd(s,d)=1$.
When $m=\gcd(s,d)>1$ the sum degenerates: $|\hat g(p)|$ equals
$\sqrt{md}$ on the sublattice $p\in m\mathbb Z_d$ (or a coset of it)
and vanishes elsewhere, so flatness fails maximally --- e.g.\ at $d=8$,
$s=2$ the spectrum is $(4,0,4,0,4,0,4,0)$, and at $d=12$, $s=4$ it is
$(0,0,6.93,0,\dots)$. For $\theta$ off the lattice $\pi\mathbb Z/d$,
$\lambda$ is not a $2d$-th root of unity, no re-indexing is available,
and flatness fails; a fine scan of $\theta\in[0,2\pi)$ at $10^{5}$
points per dimension for $d=3,\dots,8$ finds no flat angle outside
$\{\pi s/d\}$.

The flat set is therefore exactly
\begin{equation}
  \Theta_d=\{\,\pi s/d \;:\; \gcd(s,d)=1,\ sd\ \text{even}\,\},
\end{equation}
which has $\varphi(d)$ members in $[0,2\pi)$ for odd $d$ and
$2\varphi(d)$ for even $d$ ($\varphi$ Euler's totient): two members for
$d=2,3$, four for $d=4,5$, eight for $d=8$. Every member works
identically in the protocol, with the decoder modified as stated after
Eq.~\eqref{eq:lambdas} ($\lambda\mapsto\lambda_s$ in $\Lambda$,
$P_{\theta^\ast}\mapsto P_\theta$ and $R\ket u=\ket{s^{-1}u}$ in
$\Gamma$); we confirmed exact recovery on every flat angle for
$d\le9$; $\theta^{\ast}(d)$ of
Eq.~\eqref{eq:critical} is simply the smallest positive element,
$s=1$ for even $d$ and $s=2$ for odd $d$. This also explains why no
single parity-independent formula exists: $s=1$ is disallowed for odd
$d$ because $sd$ is then odd, which is precisely the failure of $2$ to
be invertible seen in Methods.

\subsection{Restriction I: qudit encrypted cloning}
\label{si:cloning}
Removing the data qudits and the teleportation stage from the protocol
leaves encrypted cloning of a single qudit, the circuit of
Fig.~\ref{fig:full_circuit}b. Here we collect what the main text and
Methods establish about this restriction on its own. The input is the
provider's register $B_C$, which may be entangled with an external
system $E$ of any dimension, so that the joint state is an arbitrary
$\ket\psi_{B_CE}$. In the full protocol $E$ is Alice's qudit $B_A$
together with anything $B_A$ is entangled with. The encoder
Eq.~\eqref{eq:encoder} acts on $(B_C,S_1,\dots,S_n)$ with each $S_j$
initially Bell-paired with a key qudit $K_j$, and the decoder
Eq.~\eqref{eq:decoder} acts on $(S_B,K_1,\dots,K_n)$.

\emph{Privacy holds at every angle.} Methods, Eq.~\eqref{eq:coalition},
shows that any set of shares short of all $n$ is left in
$\id/d^{|\mathcal C|}\otimes\rho_E$. That derivation uses only the Weyl
expansion Eq.~\eqref{eq:weylsum} and the normalization
$\sum_{p,q}|\mu_{p,q}|^{2}=d^{4}$, both valid for any $\theta$, so a
losing site learns nothing whether or not the angle is tuned. This
requires $n\ge2$. A single share has no spectator pair to suppress the
cross terms, so for $n=1$ it is the all-shares coalition of
Section~\ref{si:coalition} and is private only when $\gcd(1-\eta^{2},d)=1$; at $d=2$,
for example, its trace distance from $\id/d\otimes\rho_E$ is $1/2$,
even though recovery still works.

\emph{Recovery holds on the flat angles.} For
$\theta\in\Theta_d$ (Section~\ref{si:angle}), and with
the decoder modified as stated after Eq.~\eqref{eq:lambdas}, the
decoder at any site $B$ returns
\begin{equation}
  \ket\psi_{S_BE}\otimes\ket\Phi_{B_CK_B}\otimes
  \bigotimes_{j\ne B}\ket\Phi_{S_jK_j},
  \label{eq:si_clone}
\end{equation}
which is the equality drawn in Fig.~\ref{fig:full_circuit}b. The
Methods derivation (Steps 1--5) treats the case where $B_C$ is maximally
entangled with a reference $B'$, and it uses nothing beyond the chirp
form Eq.~\eqref{eq:chirp}. Any other input follows, because every
$\ket\psi_{B_CE}$ can be written as $\sqrt d\,(\id\otimes
V)\ket\Phi_{B_CB'}$ for some linear map $V:B'\to E$. Since the decoder
never touches $B'$, applying $V$ afterwards commutes with it and turns
$\ket\Phi_{B'S_B}$ into $\ket\psi_{S_BE}$. Entanglement between $B_C$
and the outside world is therefore transferred to the recipient
intact, not only the local state of $B_C$.

\emph{Off the flat angles, no decoder exists.} The statement concerns
the encoder family $\Uenc(\theta)$ of Eq.~\eqref{eq:encoder}, and
excludes every decoder, not only that of Eq.~\eqref{eq:decoder}, acting
on the recipient's registers. Take $B_C$ maximally
entangled with $B'$. After exact recovery $(B',S_B)$ is in a pure
maximally entangled state, so it cannot be correlated with anything
else, in particular with $B_C$, which never reaches the recipient.
Hence the encoded state must already have $\rho_{B'B_C}=\id/d^{2}$.
Tracing all signal--key pairs out of Eq.~\eqref{eq:branches} gives the
Bell-diagonal state $\rho_{B'B_C}=\sum_{p,q}|\mu_{p,q}|^{2}d^{-4}\,
(\Wpq)_{B_C}\ketbra\Phi(\Wpq)^{\dagger}_{B_C}$, which is maximally mixed
only if all $d^{2}$ weights are equal, i.e.\ $|\hat g(p)|=\sqrt d$ for
every $p$. By Section~\ref{si:angle} this is exactly the
condition $\theta\in\Theta_d$. Numerically, at off-lattice angles the decoder fidelity drops well
below one (to $0.15$ at $d=3$ and $0.065$ at $d=4$ in our tests),
while the unselected shares remain exactly maximally mixed.

\emph{Cost.} The encoder uses $4n$ two-qudit gates, two phase gates and
two Fourier layers, and the decoder uses $3n$ two-qudit gates and three
single-qudit gates. Neither count depends on $d$.

\emph{Replacing lost keys.} Suppose the keys $K_j$ for $j$ in some set
$L$ of $\ell$ unselected sites are lost, while the corresponding shares
$S_j$ are still coherent and can be handed to the recipient. The two
halves of a Bell pair are interchangeable only up to a transpose:
$(\Wpq\otimes\id)\ket\Phi=(\id\otimes\Wpq^{T})\ket\Phi$ with
$\Wpq^{T}=\Zdop^{q}\Xd^{-p}=\omega^{-pq}W_{-p,q}$. Seen from $S_j$, the
shift label therefore appears with the opposite sign, the phase label
with the same sign, and each exchanged pair carries an extra
$\omega^{-pq}$. The decoder compensates with three changes for every
$j\in L$: $\CSUM(S_B\!\to\!K_j)$ becomes $\CSUB(S_B\!\to\!S_j)$,
$\CSUM(K_j\!\to\!K_B)$ becomes $\CSUM(S_j\!\to\!K_B)$, and the block
$\CSUB(S_B\!\to\!K_B)^{n}$ shortens to $\CSUB(S_B\!\to\!K_B)^{n-\ell}$.
Recovery is then again exact, with every pair $(S_j,K_j)$, $j\in L$,
returned to $\ket\Phi$. We confirmed this numerically for $d=3,4,5$
with one and two replaced keys. Using the unmodified decoder on the
substituted registers gives entanglement fidelity $1/d^{2}$ in general.

\emph{Relation to other work.} For $d=2$ the encoder coincides with
that of Ref.~\cite{YamaguchiKempf2026cloning} up to the global phase
$i$. Two recent works also construct qudit encrypted cloning.
Ref.~\cite{Ceara2026qudit} encodes with Chu sequences, which are
quadratic-phase sequences closely related to $P_{\theta^{\ast}}$, and
gives a decoder whose two-qudit gate count grows as $d^{3}$.
Ref.~\cite{LimLo2026AME} works in the Weyl--Heisenberg picture, relates
the construction to absolutely maximally entangled states, and gives
the decoder as an operator sum. What the present construction adds is
the explicit $3n$-gate decoder, independent of $d$, exact transfer of
entanglement with an arbitrary external system, and the statement that,
for this encoder family, the flat angles are necessary as well as
sufficient.

\subsection{Coalition privacy and the all-shares coalition}
\label{si:coalition}
Eq.~\eqref{eq:coalition} of Methods proves, in closed form and for all
$d$ and all $n\ge2$, that \emph{every} coalition $\mathcal
C\subsetneq\{1,\dots,n\}$ of signal qudits holds exactly
$\id/d^{|\mathcal C|}$, uncorrelated with the reference. This covers
the entire threat model: in the deferred-recipient setting the selected
site is chosen after distribution, so a coalition that already contains
every site has nothing left to learn about. The bound is tight at
$|\mathcal C|=n-1$ and requires no numerical support. Direct evaluation
over all subsets for $d\le7$, $n\le4$ reproduces trace distance
$<10^{-16}$ in every case, as it must.

The one case the spectator argument cannot reach is
$\mathcal C=\{1,\dots,n\}$ --- all $n$ recipients colluding \emph{and}
pooling their shares before any selection is made. This lies outside
the threat model, but it is worth recording that it is genuinely not
private in general, rather than merely unproven. With no spectator pair
to enforce orthogonality, cross terms $(p,q)\ne(p',q')$ survive the
partial trace, and the residual correlation with $B_A$ is governed by
the overlap of the pooled signal registers across branches. The
stabilizer count of Section~\ref{si:access} (case $b=0$, $s=n$, $k=0$)
gives a surviving correlation unless
\begin{equation}
  \gcd\!\left(n-\eta^{2},\,d\right)=1,\qquad
  \eta=\begin{cases}1, & d\ \text{even},\\ 2^{-1}\bmod d, & d\ \text{odd},\end{cases}
\end{equation}
so for example $(d,n)=(2,3)$ fails this criterion and the pooled
coalition recovers $B_A$-correlated information, with trace distance
$\tfrac12$ between $\rho_{B_A\mathcal C}$ and
$\rho_{B_A}\otimes\id/d^{n}$. We note two mitigating facts. First, the
data qudit $D_A$ remains perfectly protected even here: its reduced
state conditioned on the full coalition is unchanged, trace distance
$0$, because $D_A$ is folded into $B_A$ by a gate the encoder never
touches. Second, any protocol run with $\gcd(n-\eta^{2},d)=1$ is
private even against the full coalition. Choosing $n$ to satisfy this
congruence is free, since $n$ is a design parameter; for $n=2$ it
reduces to $\gcd(7,d)=1$ for odd $d$ and holds for every even $d$,
the same condition as the $((3,5))$ structure of
Section~\ref{si:access} at $\theta^{\ast}$.

\subsection{Access structure of the encrypted resource}
\label{si:access}
Here we treat the encoded register as a secret-sharing scheme. The
secret is the provider's reference $R$ --- Alice's half $B_A$, or
anything $B_C$ was entangled with --- and the $2n+1$ shareholders are
$B_C,S_1,\dots,S_n,K_1,\dots,K_n$. For a set $A$ of these registers we
ask how much of the secret it holds, measured by the mutual information
$I(R\!:\!A)$ in units of its maximum $2\log d$. The value $1$ means that
$A$ can recover $R$'s partner exactly, and $0$ means that $A$ is
uncorrelated with $R$. Because the global state is pure,
$I(R\!:\!A)+I(R\!:\!\bar A)=2\log d$ for the complement $\bar A$, so a
set is authorized exactly when its complement learns nothing. Classify
each signal--key pair by how much of it lies in $A$. Let $f$ be the
number of pairs with both members in $A$, $s$ and $k$ the numbers with
only $S_j$ or only $K_j$ in $A$, $u=n-f-s-k$ the number of pairs
entirely outside $A$, and $b=1$ if $B_C\in A$.

\emph{Sets that miss a whole pair.} If $u\ge1$, a pair outside $A$ is a
spectator exactly as in Eq.~\eqref{eq:coalition}. The cross terms of
Eq.~\eqref{eq:branches} vanish, and $A$ holds an incoherent mixture over
the labels $(p,q)$ with weights $|\mu_{p,q}|^{2}/d^{4}$. A single member
of a pair is maximally mixed for every label, so half-pairs reveal
nothing. A complete pair $(S_j,K_j)$ reveals $(p,q)$ exactly, through
the orthonormal states $(\Wpq\otimes\id)\ket\Phi$. With $B_C$ and one
complete pair, the label can be read out and $\Wpq^{\dagger}$ undone on
$B_C$, restoring $\ket\Phi_{RB_C}$, so $I=1$. With $B_C$ but no complete
pair, $B_C$ is Weyl-twirled, and at a flat angle the twirl is uniform,
leaving $R B_C$ in $\id/d^{2}$, so $I=0$. Without $B_C$, $A$ never
touches $R$'s partner, so $I=0$ again. Hence, for $u\ge1$,
\begin{equation}
  I(R\!:\!A)=1\ \text{if}\ b=1\ \text{and}\ f\ge1,\quad
  I(R\!:\!A)=0\ \text{otherwise}.
  \label{eq:access_spectator}
\end{equation}
Only the step with $B_C$ and no complete pair uses the flat angle. Off
the lattice the twirl is not uniform, and $B_C$ keeps partial
information.

\emph{Sets that touch every pair.} If $u=0$ and $f\ge1$, the complement
contains no complete pair and misses pair $f$'s partner entirely, so by
Eq.~\eqref{eq:access_spectator} it learns nothing, and $A$ is
authorized. The remaining sets hold exactly one member of each pair,
with or without $B_C$. Here no spectator exists, and the answer
depends on arithmetic. At the flat angle $\theta=\pi r/d$ it is
\begin{equation}
  I(R\!:\!A)=
  \begin{cases}
   1-\dfrac{\log g_{1}}{2\log d}, & b=1,\\[8pt]
   \dfrac{\log g_{0}}{2\log d}, & b=0,
  \end{cases}
  \label{eq:access_gcd}
\end{equation}
with $g_{1}=\gcd\!\big(r^{2}k(s+1)-1,\,d\big)$ and
$g_{0}=\gcd\!\big(r^{2}s(k+1)-1,\,d\big)$,
for $s+k=n$. The two cases are complements of each other and sum to
$1$, as purity requires. The case $b=0$, $s=n$, $k=0$ is the pooled
coalition of all $n$ shares of Section~\ref{si:coalition}, and
$g_{0}=\gcd(r^{2}n-1,d)$ reproduces its condition
$\gcd(n-\eta^{2},d)=1$ (for odd $d$, $r=2$ and $4n-1=4(n-\eta^{2})$
with $4$ invertible; for even $d$, $r=1$ and $\eta=1$).

{\emph{Relation to earlier classifications.} For qubits and sets
without $B_C$, $g_0$ reproduces the parity rule of
Ref.~\cite{Gianini2026leak}: a set with $s$ signal and $k$ key qubits
leaks exactly when $s$ and $n$ are both odd. Sets containing the
source were classified for qubits in Ref.~\cite{Gianini2026full}. For
even $d$ ($r=1$), $g_0=\gcd(s(k+1)-1,d)$ is the leakage criterion of
Ref.~\cite{Bai2026classification}, derived there for sets without the
source; for odd $d$ that work uses a different quadratic phase, and
the factor $r^{2}$ reflects the encoder of Eq.~\eqref{eq:encoder}. New
here are the case $b=1$ for $d>2$ and the derivation of both cases,
together with Eq.~\eqref{eq:access_spectator}, from a single
stabilizer count.

\emph{Derivation.} Both Eq.~\eqref{eq:access_spectator} and
Eq.~\eqref{eq:access_gcd} follow from a single counting argument,
because at a flat angle the encoder is a Clifford unitary and the
encoded register is a stabilizer state. Let $T=\sum_c t_c$ be the sum
over the $N=n+1$ carriers $(B_C,S_1,\dots,S_n)$; because $rd$ is even,
$e^{i\theta T^{2}}$ depends only on $T$ modulo $d$, and
$e^{i\theta[(T+1)^{2}-T^{2}]}=e^{i\theta}\omega^{rT}$. Hence $U_Z$ maps each
carrier shift $X_c$ to $X_cZ^{r}_{\rm all}$, up to phase, and fixes
every $Z_c$; conjugating by $\mathbb F$ gives the action of $U_X$. On
a carrier Weyl string labeled by $(\mathbf x,\mathbf z)\in\Zd^{2N}$,
$\Uenc$ therefore acts as the linear map
\begin{equation}
  \mathbf x\mapsto\mathbf x-\big(r^{2}N\sigma_x+r\sigma_z\big)\mathbf 1,
  \qquad
  \mathbf z\mapsto\mathbf z+r\sigma_x\mathbf 1,
  \label{eq:symplectic}
\end{equation}
with $\sigma_x=\sum_cx_c$, $\sigma_z=\sum_cz_c$ and $\mathbf 1$ the
all-ones vector. Before encoding, the $2N$ qudits $R,B_C,\{S_j,K_j\}$
are in Bell pairs, stabilized by $X\!\otimes\!X$ and
$Z\!\otimes\!Z^{-1}$ on each pair. A general stabilizer element is
labeled by $(a,\beta,a_j,b_j)\in\Zd^{2N}$, and after
Eq.~\eqref{eq:symplectic} it acts as
\begin{equation}
\begin{aligned}
  R&:(a,\beta), & B_C&:(a-\gamma,\,-\beta+r\alpha),\\
  K_j&:(a_j,-b_j), & S_j&:(a_j-\gamma,\,b_j+r\alpha),
\end{aligned}
\end{equation}
where a pair $(x,z)$ denotes $X^{x}Z^{z}$ up to phase,
$\alpha=a+\sum_ja_j$ and $\gamma=r^{2}N\alpha+r(-\beta+\sum_jb_j)$.
For a stabilizer state on $m$ qudits, the reduced state on any subset
$A$ is $\rho_A=d^{-|A|}\sum_{g\in\mathcal S_A}g$, where $\mathcal S_A$
is the subgroup of stabilizers supported on $A$, so
$S(A)=|A|\log d-\log|\mathcal S_A|$ for every $d$, prime or
not~\cite{Gheorghiu2014standard}. Since $R$
alone is maximally mixed, this gives
\begin{equation}
  I(R\!:\!A)=\log\frac{|\mathcal S_{RA}|}{|\mathcal S_A|}
  =\log\big|\pi_R(\mathcal S_{RA})\big|,
  \label{eq:MIcount}
\end{equation}
the number of values $(a,\beta)\in\Zd^{2}$ that occur on $R$ among
stabilizers vanishing outside $RA$. Computing $I$ is thus a matter of
solving linear congruences.

For a set with one member of each pair, every half-pair outside $A$
imposes vanishing conditions: if only $S_j\in A$, then $a_j=b_j=0$; if
only $K_j\in A$, then $a_j=\gamma$ and $b_j=-r\alpha$. Substituting,
$\alpha=a+k\gamma$ and $\gamma=r^{2}(s+1)\alpha-r\beta$. If $B_C\in A$,
eliminating $\alpha$ leaves the single condition
$\big(r^{2}k(s+1)-1\big)\gamma\equiv r\big[\beta-r(s+1)a\big]$. Since
$r$ is invertible modulo $d$, it is solvable for exactly $d^{2}/g_1$
pairs $(a,\beta)$, which gives the first line of
Eq.~\eqref{eq:access_gcd}. If $B_C\notin A$, its part must vanish as
well, $a=\gamma$ and $\beta=r\alpha$, which forces
$\beta=r(k+1)a$ and $\big(r^{2}s(k+1)-1\big)a\equiv0$. That has
$g_0$ solutions, giving the second line. If instead some pair lies
entirely outside $A$, both of its members must vanish, which forces
$\alpha=\gamma=0$. Then $(a,\beta)$ ranges over all of $\Zd^{2}$ when
$B_C\in A$ and $A$ contains a complete pair, whose free $(a_j,b_j)$
absorb the constraints, and only over $(0,0)$ otherwise. This is
Eq.~\eqref{eq:access_spectator}. The flat angle enters only through
Eq.~\eqref{eq:symplectic}: off the lattice $\Uenc$ is not Clifford
and the counting does not apply, consistent with the partial
information retained by $B_C$ there.

\emph{Consequences.} (i) The key register alone ($k=n$, $b=0$, $s=0$)
has $g_{0}=\gcd(-1,d)=1$ and carries no information for any $d$ and
$n$. Intercepting the keys alone, or discarding them, reveals nothing.
(ii) The recipient's set $(S_B,K_1,\dots,K_n)$ has $f=1$ and $u=0$ and
is authorized without $B_C$, which is why the provider may discard
$B_C$ after encoding. (iii) For $n=2$ there are five shareholders. By
Eq.~\eqref{eq:access_spectator}, every pair of registers carries
nothing except possibly the half-pair sets ($S_1S_2$, $S_1K_2$,
$K_1S_2$, $K_1K_2$), and every triple is authorized except possibly
$B_C$ plus two half-pairs. By Eq.~\eqref{eq:access_gcd}, these
exceptions have $(s,k)\in\{(2,0),(1,1),(0,2)\}$, and every one of them
reduces to either $\gcd(-1,d)=1$ or $\gcd(2r^{2}-1,d)$, so all of them
vanish exactly when $\gcd(2r^{2}-1,d)=1$. The scheme is then
a $((3,5))$ threshold scheme. That encrypted cloning with two pairs
is equivalent to a $((3,5))$ scheme in every dimension was shown by
Lim and Lo~\cite{LimLo2026AME} (their Theorem~3); what the present
count adds is the exact condition for the encoder of
Eq.~\eqref{eq:encoder}, which fails at $d=7$, $\theta^{\ast}$, and the
structure for $n\ge3$ below. At $\theta^{\ast}$ ($r=2$ for odd $d$, $r=1$
for even $d$) the condition reads $\gcd(7,d)=1$ for odd $d$ and holds
automatically for even $d$. For $d=7$ another flat angle, for example
$r=4$, satisfies it, and we find such an angle for every $d<3000$.
(iv) For $n\ge3$ the structure is no longer a threshold one. Sets
of one member per pair can carry exactly half of the information, for
example the three shares at $(d,n)=(2,3)$, as already noted in
Section~\ref{si:coalition}.

\emph{Verification.} As an independent check on the derivation, the
verification notebook (Code availability) computes $I(R\!:\!A)$ from
the exact state vector for every proper subset $A$ and compares it
with Eqs.~\eqref{eq:access_spectator} and~\eqref{eq:access_gcd}. It covers $d=2,\dots,9$ at $n=2$,
$d=2,\dots,5$ at $n=3$, $(d,n)=(2,4),(3,4),(2,5)$ at $\theta^{\ast}$,
and six further flat angles, 4\,182 subsets in all. The largest
deviation is $\lesssim10^{-14}$.

\begin{figure*}[t]
\centering
\figbox{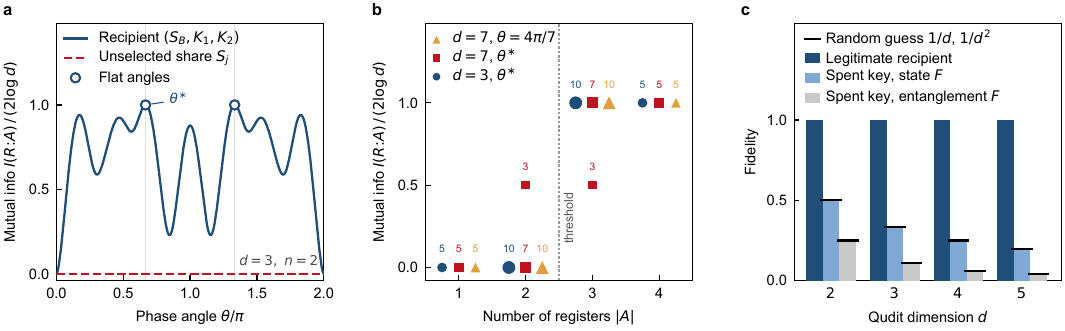}{\textwidth}
\caption{\textbf{Privacy, recovery and the access structure of the shared gate.}
\textbf{a}, Mutual information $I(R\!:\!A)$ with the reference $R$, in
units of its maximum $2\log d$, as a function of the encoder's phase
angle $\theta$, for a single unselected share ($A=S_j$, dashed) and for
the recipient's set ($A=\{S_B,K_1,K_2\}$, solid), at $d=3$, $n=2$. The
share carries nothing at every angle; the recipient recovers the
reference exactly only on the flat angles (open circles), the smallest
of which is $\theta^{\ast}$. \textbf{b}, $I(R\!:\!A)$ for every subset
$A$ of the five registers $\{B_C,S_1,S_2,K_1,K_2\}$ ($n=2$), grouped by
size $|A|$; numbers give how many subsets share each value, and marker
area scales with that count. At $d=3$, $\theta^{\ast}$ (circles) and at
$d=7$, $\theta=4\pi/7$ (triangles) every set of at most two registers
holds nothing and every set of at least three holds everything: a
$((3,5))$ threshold scheme. At $d=7$, $\theta^{\ast}$ (squares) six
sets hold exactly half, as predicted by Eq.~\eqref{eq:access_gcd}.
\textbf{c}, Fidelity obtained by a second site attempting to decrypt
with the spent key: state fidelity $1/d$ and entanglement fidelity
$1/d^{2}$, both the values of a random guess (black bars), against
fidelity exactly $1$ for the legitimate recipient.}
\label{fig:privacy}
\end{figure*}

\subsection{Illustration of privacy, recovery and access structure}
\label{si:privacyfig}
The statements illustrated in Fig.~\ref{fig:privacy} are proved in
closed form: privacy of every proper coalition by the Weyl twirl,
Eq.~\eqref{eq:coalition}; recovery on the flat angles by the chirp
identity, Eq.~\eqref{eq:chirp}, and the explicit decoder; the access
structure by Eqs.~\eqref{eq:access_spectator} and~\eqref{eq:access_gcd};
and the spent-key fidelities by the post-decoding product form,
Eq.~\eqref{eq:postdecode}. The figure makes the separation between
privacy and recovery visible. A single unselected share holds no
information about the reference at any phase angle, whereas the
recipient's set recovers the reference completely only on the flat
lattice and partially elsewhere (panel~\textbf{a}). The tuning of
$\theta$ therefore buys recovery, not privacy. Panel~\textbf{b} shows
every subset of the five registers for $n=2$. At $(d,\theta)=(3,\theta^{\ast})$
the information jumps from $0$ to $1$ between two and three registers,
which is the $((3,5))$ threshold structure. At $d=7$, where
$\gcd(2r^{2}-1,d)=7$ at $\theta^{\ast}$, three two-register sets and
three three-register sets carry exactly half the information, and
the flat angle $4\pi/7$ restores the threshold.

\subsection{Restriction II: deterministic non-local \texorpdfstring{$\CSUM^{k}$}{CSUM^k}}
\label{si:teleport}
Removing the cloning layer and fixing the recipient in advance leaves
the teleportation stage of Fig.~\ref{fig:full_circuit}c, in which the
resource pair $(B_A,B_B)$ is supplied directly. $B_B$ plays the role of
the recovered share $S_B$ of the full protocol. This is the qudit
version of the qubit non-local
$\mathtt{CNOT}$~\cite{Eisert2000,GottesmanChuang1999teleport}, and it
extends to every power of $\CSUM$.

Let Alice apply $\CSUM^{k_1}(D_A\!\to\!B_A)$ and Bob apply
$\CSUB^{k_2}(B_B\!\to\!D_B)$, for any $k_1,k_2\in\Zd$. Alice measures
$B_A$ in the computational basis ($m_A$). Bob applies $F$ to $B_B$ and
measures it in the computational basis ($m_B$). They exchange the two
outcomes, and Bob applies $\Xd^{k_2m_A}$ to $D_B$ while Alice applies
$\Zdop^{k_1m_B}$ to $D_A$. Following a basis input $\ket{c,t}$ as in
Methods, each branch $(m_A,m_B)$ occurs with probability $1/d^{2}$ and
leaves
\begin{equation}
  \ket{c}_{D_A}\ket{t+kc}_{D_B},\qquad k\equiv k_1k_2\pmod d,
\end{equation}
with the overall phase $\omega^{m_Am_B}$. That phase does not depend on
$c$ or $t$. After correction, each branch therefore applies a fixed
unitary, $\CSUM^{k}$, up to a global phase. By linearity the same then
holds for superposed inputs and for inputs entangled with external
systems, with no averaging over branches and nothing discarded.

Several choices of $(k_1,k_2)$ are worth singling out. Panel c of
Fig.~\ref{fig:full_circuit} is $k_1=k_2=1$, which gives $\CSUM$
with corrections $\Xd^{m_A}$ and $\Zdop^{m_B}$, the teleportation stage
of the main protocol. The naive transcription of the qubit circuit,
in which Bob's gate adds ($k_1=1$, $k_2=-1$), is not wrong so much as
mislabelled: it teleports $\CSUB$
exactly, provided Bob applies $\Xd^{-m_A}$. It fails only when combined
with the $\CSUM$ correction $\Xd^{m_A}$, and then it succeeds on just a
fraction $\nu(d)/d$ of (input, branch) pairs (Section~\ref{si:asymmetry}). Finally,
Bob's operations depend on $k_2$ alone. If Bob always uses $k_2=1$,
Alice selects $k=k_1$, and hence which power of $\CSUM$ is applied.
Bob's operations and the public outcomes $(m_A,m_B)$, which are
uniformly distributed for every $k$, are then independent of $k$, so
Bob learns about $k$ only what the action of $\CSUM^{k}$ on his own
qudit $D_B$ reveals, as it would for an ideal local implementation of
the gate (Methods, ``Hiding the power of $\CSUM$''). All of this carries over unchanged to the
full deferred-recipient protocol, whose final stage is exactly this
circuit acting on the recovered pair $(B_A,S_B)$.

The resources are one maximally entangled qudit pair and two classical
dits, one in each direction, as in the qubit case~\cite{Eisert2000}.
For $\gcd(k,d)=1$ the entanglement cost is optimal. In that case
$\CSUM^{k}=(M_k^{-1}\otimes\id)\,\CSUM\,(M_k\otimes\id)$ with
$M_k\ket c=\ket{kc}$, so the gate turns the product state
$F\ket0\otimes\ket0$ into a maximally entangled state of Schmidt rank
$d$. Local operations and classical communication cannot create that
entanglement, so at least one maximally entangled pair must be
consumed. For general $k$, the operator Schmidt rank of
$\CSUM^{k}=\sum_c\ketbra{c}\otimes\Xd^{kc}$ is $d/\gcd(k,d)$, and $k=0$
is the identity. Because the gate carries information in both
directions ($D_A\to D_B$ in the computational basis and $D_B\to D_A$ in
the Fourier basis), classical communication is needed both ways. We
have not shown that one dit in each direction is the minimum for every
$d$ and $k$.

\subsection{The literal qubit transcription}
\label{si:asymmetry}
By Eq.~\eqref{eq:csumk} the teleported gate is fixed by the product
$k_1k_2$, so $\CSUM$ requires the two entangling gates to have opposite
signs; whether Alice or Bob carries the inverse is immaterial, provided
the corrections are chosen accordingly. Here we quantify what happens
when the qubit circuit is transcribed literally: both parties apply
$\CSUM$ ($k_1=1$, $k_2=-1$), and the qubit corrections $\Xd^{m_A}$,
$\Zdop^{m_B}$ are kept. This circuit teleports $\CSUB$, and after the
correction $\Xd^{m_A}$ the target $D_B$ is left displaced by
$2(m_A-c)$, where $c$ is the (generally superposed) value of $D_A$. No
classical function of $(m_A,m_B)$ applied by Bob can supply this
$c$-dependent shift without measuring $D_A$; the shift is removed only
by changing Alice's side as well (her gate to $\CSUB$ and her
correction to $\Zdop^{-m_B}$), i.e.\ by restoring $k_1k_2\equiv1$. At
$d=2$ the displacement $2(m_A-c)\equiv0$ for every $c$, which is why
the sign is invisible for qubits.

The displacement vanishes exactly on the solutions of
$2(m_A-c)\equiv0\pmod d$. For odd $d$, $2$ is invertible and each $c$
is matched by the single branch $m_A=c$. For even $d$ the congruence is
$m_A\equiv c\pmod{d/2}$, solved by two values of $m_A$ for each $c$.
Either way the transcribed circuit is exact on a fraction
\begin{equation}
  \frac{\#\{(c,m_A):2(m_A-c)\equiv0\}}{d^{2}}=\frac{\nu(d)}{d},
  \qquad \nu(d)=\gcd(2,d),
  \label{eq:uninvfrac}
\end{equation}
of the $d^{2}$ (input, branch) pairs --- all of them at $d=2$, $1/3$ at
$d=3$ (Fig.~\ref{fig:obstruct}c), $1/2$ at $d=4$, $1/5$ at $d=5$, and
$\to0$ as $d$ grows --- with infidelity $1.000$ on every remaining pair,
to be contrasted with $<10^{-15}$ on all $d^{2}$ pairs when
$k_1k_2\equiv1$. The surviving successes carry no
operational content: whether a given branch is one of them is fixed by
$c$, which is precisely the unknown the protocol must not measure, so
the parties can neither predict nor post-select them. The failure is a
statement about $\mathbb Z_d$ arithmetic, not a numerical finding:
$\CSUM$ is its own inverse on $\mathbb Z_2$ and on no other
$\mathbb Z_d$. It is a pitfall in carrying qubit circuits over to
qudits rather than an obstruction, and the protocol avoids it by
construction. Supplementary
Section~\ref{si:teleport} shows the transcribed circuit is nonetheless
an exact protocol for $\CSUB$, with the correction $\Xd^{-m_A}$.

\subsection{Comparison with simpler deferral schemes, and threat model}
\label{si:context}
The introduction notes that the obvious ways of deferring the recipient
each fail one of the requirements met by our construction. Table~\ref{tab:compare} summarizes the comparison; here we spell out
the two simplest alternatives, comment on operation sharing and on the
cost of selection, and then state the scope of our guarantees.

\begin{table*}[t]
\caption{\textbf{Ways to defer or share a non-local gate.} Columns give
the resource distributed in advance, the quantum communication needed
once the recipient is chosen, who can block the gate, and who can
complete it alone without the selected node. QOS: quantum operation
sharing~\cite{Zhang2011shared,Wang2007controlled}; CT: threshold-controlled
teleportation~\cite{SaiToh2009threshold}.}
\label{tab:compare}
\begin{ruledtabular}
\footnotesize
\begin{tabular}{p{3.2cm}p{4.0cm}p{3.6cm}p{2.9cm}p{2.6cm}}
Scheme & Distributed in advance & Quantum comm.\ at selection & Can block & Can use alone\\
\hline
Entanglement swapping & $n+1$ Bell pairs to provider & none (2 dits) & provider & provider (holds $B_C$)\\
GHZ splitting & $(n+1)$-qudit GHZ state & none ($n-1$ announced dits) & any unselected node & nobody\\
QOS / CT & multipartite state, recipient fixed & none (all controllers announce) & any controller$^{a}$ & nobody\\
This work & Bell pair $+$ $n$ encrypted shares & $n$ key qudits to recipient & key custodians only & nobody\\
This work, keys pre-teleported & as above $+$ $n^{2}$ custodian--node Bell pairs & none ($2n$ dits) & key custodians only & nobody\\
\end{tabular}
\end{ruledtabular}
\raggedright\footnotesize{$^{a}$In $(k,m)$-threshold variants, any $m-k+1$ controllers.}
\end{table*}

\emph{Entanglement swapping.} The provider shares a Bell pair with each
candidate, keeps $B_C$, and once the recipient is known swaps $B_C$
onto that candidate's pair with one Bell measurement and two classical
dits. Unselected candidates learn nothing, and the decision costs two
dits rather than $n$ key qudits. But whoever holds $B_C$ holds half of
a Bell pair with Alice and can complete the gate directly: the right to
choose the recipient cannot be separated from the ability to use the
gate, and the resource is lost with $B_C$. In our construction the key
register carries no information about Alice's qudit, nor does any
single share or $B_C$, and $B_C$ may be discarded after encoding.

\emph{GHZ-based information splitting.} The provider prepares a
$(n+1)$-qudit GHZ state $d^{-1/2}\sum_j\ket{j}_{B_A}\ket{j}^{\otimes n}$
over $B_A$ and the candidates~\cite{Hillery1999sharing,Karlsson1999sharing}.
To leave a Bell pair between Alice and a chosen site, each of the
$n-1$ other sites must measure in the Fourier basis and announce its
outcome for a phase correction. Every unselected site must therefore
participate, any one of them can block or corrupt the gate by
withholding or misreporting its outcome, and all of them learn that a
selection has been made. Moreover each GHZ share is classically
correlated with $B_A$ in the computational basis: measuring one share
reveals the outcome Alice would obtain. In our construction
unselected sites take no action, each share is exactly maximally mixed
and uncorrelated with $B_A$, and they cannot tell whether a selection
has occurred.

\emph{Operation sharing and controlled teleportation.} In quantum
operation sharing~\cite{Zhang2011shared,Wang2007controlled} and
threshold-controlled teleportation~\cite{SaiToh2009threshold} a dealer
enables an operation on, or a teleportation to, a recipient fixed in
advance, and the sharers are controllers who must each measure and
announce. These schemes therefore share the participation, veto and
visibility properties of GHZ splitting. Here the recipient is one of
the sharers, chosen after distribution, the unselected sharers take no
action, and the shared object is an entangling gate acting on data
that never enter the scheme.
}

\emph{Cost of selection.} Swapping is faster at selection: it moves
two dits, whereas our construction moves $n$ key qudits to the
recipient. This cost can be moved to the distribution stage by
pre-sharing $n$ Bell pairs between the key custodian and each candidate
($n^{2}$ pairs) and teleporting the keys at selection with $2n$ dits,
after which selection is classical, as for swapping. The advantage of
our construction is therefore not raw selection latency but the
distribution of trust: the provider holds nothing usable, authority can
be split among custodians, and unselected nodes neither act nor learn
of the selection.

\emph{Scope and threat model.} Apart from the noise analysis, our results assume perfect state
preparation, gates and measurements and lossless classical
communication, and $n\ge2$ candidate sites with $2n+4$ qudits in total
(for $n=1$ recovery remains exact, but the single share is private
only when $\gcd(1-\eta^{2},d)=1$; Section~\ref{si:coalition}). The privacy guarantee covers honest-but-non-participating
parties---sites that observe without deviating from the protocol---and
not active corruption of a site's operations. The provider is trusted
to prepare the specified states and to release the keys to one site; it
and Alice know the recipient's identity, and anonymity is claimed only
toward the unselected sites, at the level of their registers and the
public transcript. The mere presence of a classical exchange between
Alice and some site is not hidden by the protocol. If that matters,
the dits can be sent privately, or uniformly random dits can be
published in rounds without a selection, which is indistinguishable
because the genuine $(m_A,m_B)$ are uniform. The two operational
requirements stated in the Protocol section mark the boundary
explicitly. (i) Key delivery: the key register alone carries nothing,
but an unselected site that intercepts all of it holds an authorized
set together with its own share and could redeem the gate in Bob's
place, so the keys must travel over an authenticated quantum
channel~\cite{Barnum2002authentication}. (ii) Classical messages:
$m_A$ and $m_B$ may be public but must be authenticated. An adversary
that alters them by $(\delta_A,\delta_B)$ learns nothing and can only
append the Weyl error $\Zdop^{k_1\delta_B}\otimes\Xd^{k_2\delta_A}$ of
Eq.~\eqref{eq:tamper} to the gate. When the
keys are held by independent custodians, as discussed in the main text,
each custodian's key is by itself uncorrelated with Alice's qudit
(Section~\ref{si:access}).

\end{document}